\documentclass[a4paper,fleqn,usenatbib]{mnras}

\usepackage[T1]{fontenc}
\usepackage{ae,aecompl}

\usepackage{graphicx}	
\usepackage{amsmath}	
\usepackage{amssymb}	

\title[H {\sc i} enhancement in proto-cluster]{Enhancement of H~{\sc i} absorption associated with the $z=3.1$ 
large-scale proto-cluster and characteristic structures with AGNs sculptured over Gpc scale in the SSA22 field} 

\author[T.\ Hayashino et al.]{\parbox[t]{\textwidth}{\vspace{-1cm}
T. Hayashino$^{1}$\thanks{E-mail: haya@awa.tohoku.ac.jp, reston@jcom.home.ne.jp}, 
A. K. Inoue$^{2,3}$, 
K. Kousai$^{1}$, 
N. Kashikawa$^{4,5,6}$, 
K. Mawatari$^{2,7}$, 
Y. Matsuda$^{4,5}$, 
N. Tejos$^{8,3}$, 
J. X. Prochaska$^{3,9}$,  
I. Iwata$^{5,10,11}$, 
S. Noll$^{12,13,14}$, 
D. Burgarella$^{15}$, 
T. Yamada$^{16,17}$, 
M. Akiyama$^{16}$} 
\\\\
$^{1}$Research Center for Neutrino Science, Graduate School of Science,
Tohoku University, Aramaki, Aoba-ku, Sendai 980-8578, Japan\\
$^{2}$Department of Environmental Science, Faculty of Design Technology, Osaka Sangyo University, 
3-1-1, Nakagaito, Daito, Osaka \\ 574-8530, Japan\\
$^{3}$Department of Astronomy and Astrophysics, University of
California Santa Cruz, 1156 High Street, Santa Cruz, CA 95064, USA\\
$^{4}$National Astronomical Observatory of Japan, 
2-21-1, Osawa, Mitaka, Tokyo 181-8588, Japan\\
$^{5}$The Graduate University for Advanced Studies (SOKENDAI), Tokyo
181-8588\\
$^{6}$Department of Astronomy, Graduate School of Science, 
The University of Tokyo, 7-3-1 Hongo, Bunkyo, Tokyo 113-0033, Japan\\
$^{7}$Institute for Cosmic Ray Research, The University of Tokyo, 
5-1-5 Kashiwanoha, Kashiwa, Chiba 277-8582, Japan\\
$^{8}$Instituto de F\'isica, Pontificia Universidad Cat\'olica de Valpara\'iso, 
Casilla 4059, Valpara\'iso, Chile\\
$^{9}$University of California Observatories, Lick Observatory 1156 High
Street, Santa Cruz, CA 95064, USA\\
$^{10}$Subaru Telescope, National Astronomical Observatory of Japan, 
650 North A'ohoku Place Hilo, HI 96720, USA\\
$^{11}$Department of Astronomy and Physics and Institute for
Computational Astrophysics, Saint Mary's University, 923 Robie Street,\\
 Halifax, Nova Scotia B3H 3C3, Canada\\
$^{12}$Institut f\"ur Astro- und Teilchenphysik, Universit\"at
Innsbruck, Technikerstr.\ 25/8, 6020 Innsbruck, Austria\\
$^{13}$Institut f\"ur Physik, Universit\"at Augsburg, Universit\"atsstr. 1, 
86159 Augsburg, Germany\\ 
$^{14}$Deutsches Fernerkundungsdatenzentrum, Deutsches Zentrum f\"ur 
Luft- und Raumfahrt, M\"unchener Str. 20, 82234 \\
We\ss{}ling-Oberpfaffenhofen, Germany\\ 
$^{15}$Laboratoire d'Astrophysique de Marseille, Observatoire
Astronomique de Marseille-Provence, 38 rue Fr\'ed\'eric Joliot-Curie, 13388 
\\ Marseille Cedex 13, France\\
$^{16}$Astronomical Institute, Graduate School of Science, Tohoku 
University, Aramaki, Aoba-ku, Sendai 980-8578, Japan\\
$^{17}$Institute of Space and Astronautical Science, Japan Aerospace Exploration Agency, 
Kanagawa 252-5210, Japan\\ 
}

\begin{document}

\date{}

\pagerange{\pageref{firstpage}--\pageref{lastpage}} \pubyear{2018}

\maketitle

\label{firstpage}

\begin{abstract}
In the SSA22 field which exhibits a large-scale proto-cluster at $z=3.1$, we carried out a  
spectroscopic survey for Lyman Break Galaxies (LBGs) with the VLT/VIMOS and 
identified 78 confident LBGs at $z=2.5$--4. We stacked their spectra in the observer's frame 
by using a sophisticated method. 
Analyzing the composite spectrum, we have revealed that 
the large-scale proto-cluster at $z=3.1$ has a strong H~{\sc i} absorption dip 
of rest-frame equivalent width of $-1.7$~\AA. Another strong absorption dip 
found at $z=3.28$ is associated with a modestly high-density LBG peak, similar to that at $z=3.1$. 
We have also detected an H~{\sc i} transparency peak at $z=2.98$ in the composite 
spectrum, coincident with a void in the LBG distribution. 
In this paper, we also investigated the relation between LBGs, 
H~{\sc i} gas and AGNs at $z=3$--4 in the SSA22 field. 
Two AGNs at $z=3.353$ and 3.801 are, respectively, associated with the LBG concentration of an overdensity 
factor $\delta_{\rm LBG}\simeq2$ in the present statistics. 
Another structure at $z=3.453$ is remarkable: 
20 comoving Mpc scale dense H~{\sc i} gas which is not associated with any apparent LBG overdensity 
but involving a pair of AGNs. 
Such structure may be a new type of the AGN-matter correlation.
If the inhomogeneous structures over a comoving Gpc scale found in this 
paper are confirmed with sufficient statistics in the future, the SSA22 field will 
become a key region to test the standard cold dark matter structure formation scenario.
\end{abstract}


\section{Introduction}

 Regions showing galaxy concentration at high redshift are important places
 to study formation and evolution of galaxies as well as cosmological structures.
 Among them, the proto-cluster (PC) at $z=3.1$ in the
 SSA22 field, initially discovered by \cite{ste98} can be called ``a
 treasure island'' of the Universe. In 1998, they found out a 
 number density peak of Lyman Break Galaxies (LBGs) with the overdensity of $\delta\sim5$ 
 at $z=3.1$, and their narrow-band (NB) imaging survey for the peak in 2000 detected  
 72 Ly$\alpha$ emitters (LAEs) as well as two gigantic Ly$\alpha$
 emitting objects so-called Ly$\alpha$ blobs (LABs) together with about a dozen  
 of Ly$\alpha$ absorbers (LAAs) in their $9' \times 9'$ survey field, SSA22a \citep{ste00}. 

 Following this, a deep NB imaging survey was carried
 out in 2002 in a field centered on the PC, SSA22a with Subaru
 Suprime-Cam (S-Cam) having a wide field of view (FoV) 
 of $35' \times 27'$ to see the environment of the PC \citep{hay04}. 
 We call the survey area SSA22 Sb1 or simply Sb1 in this paper. 
 As a result, 283 highly confident LAEs exhibiting a belt-like large-scale structure were found, 
 i.e., it is revealed that the PC by Steidel et al.\ is not
 isolated in high-redshift space, but a part of the much larger
 structure. Interestingly, ``the belt'' extends to the edge of the field
 of view, being 60 or more comoving Mpc long.
 In this survey, around 30 LABs \citep{mat04} and LAAs
 \citep{hay04} are found in the belt-like structure besides LAEs and LBGs.
 Here LABs are expected to be a progenitor of massive galaxies
 in the present Universe \citep[e.g.,][]{uch12}. 
 Thus, the belt-like large-scale structure
 found in the Sb1 field is like a factory producing various kinds of galaxies
 in the young Universe and can be called the ``large-scale proto-cluster
 (LSPC)''.   
 ``The belt'' would evolve to a filament of the present-day large-scale
 structure and the PC discovered by Steidel et al.\
 would collapse to a massive cluster of galaxies at present \citep[e.g.,][]{top18}.  
 Objects in the structure of Sb1 characterized by Ly$\alpha$ 
 emission or absorption have also been investigated so
 far in various wavelengths from sub-mm to X-ray.

 In succession, a panoramic NB imaging survey was performed to reveal the entire extent 
 of the structure seen in Sb1 as an Intensive Program of Subaru telescope in 2005 
 \citep{yam12}. The area of 7 S-Cam FoVs around the
 SSA22-Sb1 was observed in the panoramic 
 survey of 200\,Mpc$\times$100\,Mpc in comoving scale. 
 In the survey, around 1400 LAEs and 100 LABs were detected. 
 This survey reveals farther lateral extension of the LSPC. 
 A wide sky map of the LAEs is displayed in \cite{yam12}.
 By the stacking analysis of Ly$\alpha$ images in high statistics from the panoramic survey, 
 \cite{mat12} have found that LAEs have large Ly$\alpha$ emitting halos 
 extended to 60 proper kpc showing an interesting dependence, i.e., larger halos
 for higher LAE local density in Mpc scales. 
 The correlation between the Ly$\alpha$ halo size of
 individual objects and their Mpc
 scale environment is indeed remarkable, implying interactions 
 of these galaxies with neutral hydrogen atoms in that scale 
 probably controlled by the dark matter gravitation, 
 and suggesting the importance of studies on H~{\sc i} gas 
 in the LSPC.
 Also, by stacking Ly$\alpha$ images of LBGs in their large sample, 
 \cite{ste11} confirmed extended Ly$\alpha$ halos around LBGs,  
 which had been first detected in naive form in the SSA22 PC 
 at $z=3.1$ by \cite{hay04}.
 
 In the panoramic survey area of the 200\,Mpc$\times$100\,Mpc
 comoving scale, the PC found by 
 \cite{ste98} in Sb1 is still the highest density peak and two LABs discovered 
 by them are the biggest two of all LABs found in the panoramic field. 
 So, the Sb1 field including the original PC would be
 the most important region to be intensively studied. 
 Indeed, successive spectroscopic observations of LBGs, LAEs
 and LABs in Sb1 have been performed
 \citep{mat05,mat06,kou11,yam12b,sae15,top16} 
 and the three-dimensional structure of ``the belt'' and the PC have been
 discussed \citep{mat05,kou11,top18}. 

 It is fundamentally important to measure H~{\sc i} abundance of such structure
 to understand galaxy formation in a PC with high galaxy 
 density, because many kinds of objects in the structure characterized by
 strong Ly$\alpha$ emission and/or absorption form and evolve
 by using neutral hydrogen that may be supplied from the structure.
 From this viewpoint, a dense H~{\sc i} region associated with a high
 LBG density peak at $z=2.895$ recently discovered by \cite{cuc14} in the COSMOS field 
 in the VIMOS Ultra-Deep Survey (VUDS) is noteworthy.
 While \cite{cuc14} only probed H~{\sc i} at the PC redshift,  
 the Wiener-filtered tomographic survey in the COSMOS field by 
 \cite{lee14a,lee14b,lee16,lee18} presents interesting and impressive
 results on the correlation of the three-dimensional LBG distribution and
 H~{\sc i} absorption map on comoving Mpc scales 
 in a wide redshift range of $2.0 < z < 2.6$.  

 In 2008, we performed yet another spectroscopic survey with VLT/VIMOS, 
 hereafter VI08, having a wide field of view, to reveal LBG distribution in foreground 
 and background of the $z=3.1$ LSPC. Namely, we tried a longitudinal, i.e., line-of-sight, 
 extension of the survey region from the narrow redshift range of $z=3.06$--3.12 
 sliced by the NB filter for LAEs to a wider redshift range around $z=3$ selected 
 by the U-dropout method for LBGs. As we will see in this paper, the VI08 survey has 
 revealed 30 or more LBGs behind the LSPC at $z=3.1$. These LBGs should have important
 information on H~{\sc i} of the LSPC in their spectra.  
 Here, LBGs are not as bright as QSO/AGNs but they have a higher comoving density. 
 So, the individual LBG spectrum may be noisy to
 obtain significant information on LSPC H~{\sc i}. However, if we stack
 these spectra in the observer's frame to improve a signal to noise
 ratio (S/N), information on H~{\sc i} gas in the LSPC imprinted will
 appear. Also, the stacked spectrum would reveal H~{\sc i} gas
 distribution in foreground and background of the LSPC,
 which we present in this paper.

 As a companion analysis, we have also tried to map the H~{\sc i}
 absorption distribution at the PC redshift $z=3.1$ by using our deep NB
 imaging data with higher S/N than the individual spectra of the VI08
 survey, which is reported in \cite{maw17}. In the NB photometric data
 of galaxies behind the $z=3.1$ PC, information on H~{\sc i} absorption
 is imprinted in the spectra. This method is especially very effective 
 to study absorption with the similar wavelength widths as the NB
 filter and to depict the two-dimensional map of H~{\sc i} gas. 
 Interestingly, an H~{\sc i} absorption excess is observed throughout
 the Sb1 area corresponding to a size larger than 50 comoving Mpc 
 \citep{maw17}. 

 Also, in the SSA22 field, Sb1, around 10 AGNs are
 already detected at redshifts between $z=3$ and 4 in the precedent
 studies \citep{leh09,sae15,mic17}. 
 Therefore, we are able to investigate H~{\sc i} {\it and} 
 LBG distributions as well as their connection with the AGNs in these redshifts, 
 which is another theme presented in this paper. 

 The following is the structure of this paper;  
 in the next section, we describe the imaging and spectroscopic data of LBGs in the 
 SSA22-Sb1 field. In section 3, we present the method of the observer's frame 
 composite to examine H~{\sc i} with high S/N. In section 4, we show the results obtained 
 from the observer's frame composite analysis and investigate a correlation between 
 the LBG distribution and H~{\sc i} absorption. 
 In section 5, we consider the over-density mass and the appearance probability of the 
 LSPC at $z=3.1$ as well as characteristic surroundings showing H~{\sc i} transparency. 
 In section 6, we discuss inhomogeneous distributions of LBGs associated with AGNs 
 at $z=3$--4 and present a dense H~{\sc i} region involving a pair AGN. 
 The final section is devoted to our conclusions. 
 In appendix, we present a catalog of the VI08 LBGs. 
 We assume the flat Universe with cosmological parameters of $H_0=70$\,km\,s$^{-1}$\,Mpc$^{-1}$, 
 $\Omega_m=0.3$, and $\Omega_\Lambda=0.7$. The magnitude unit throughout this paper is 
 the AB system. 

\section{Sample of Lyman break galaxies}

\subsection{Photometric data and color selection}

The photometric data we used are $B$, $V$, $R_c$, $i'$, and $z'$-band imaging 
taken with Subaru/S-Cam (\citealt{miy02}) and $u^*$ band imaging 
taken with Canada-France-Hawaii Telescope (CFHT)/Megacam \citep{bou03}. The observations and data
reduction for the S-Cam data are described in \cite{hay04}. The Megacam
data was downloaded from the CFHT archive and reduced by a standard
manner described in \cite{kou11} and references therein. The 1-$\sigma$
limiting magnitudes in each band are 27.8 ($u^*$), 28.2 ($B$), 28.2
($V$), 28.3 ($R_c$), 27.9 ($i'$), and 27.2 ($z'$) for a $2''$ diameter
aperture. We selected LBGs from objects detected in $R_c$ by 
the following color selection criteria 
similar to those adopted in literature \cite[e.g.,][]{ste95,yos08}: 
\begin{enumerate}
 \item $23.9 \leq R_c \leq 25.4$
 \item $(U-V)-1.8(V-R_c)\geq1.1$ 
 \item $R_c-i'\leq0.3$
\end{enumerate}
The faint magnitude limit of the criterion (i) is determined to select
LBGs bright enough to detect their continuum in the follow-up
spectroscopy.

\subsection{Spectroscopic data and redshift determination}


\begin{table*}
 \centering
 \begin{minipage}{160mm}
  \caption{Summary of VIMOS observations for the SSA22 field.}
  \begin{tabular}{lcccccl}
   \hline
   & RA (J2000) & DEC (J2000) & Observations & Exposure
   time (s) & Seeing (arcsec) & Remark \\
   \hline
   FoV-1 & 22:17:31.9 & $+$00:24:29.7 & July--October 2008 & 14,080 
		   & 0.32--1.93 & Loss of quadrant \#2 \\
   FoV-2 & 22:17:39.1 & $+$00:11:00.7 & August--October 2008 & 14,080
		   & 0.48--1.37 \\
   \hline
  \end{tabular}
 \end{minipage}
\end{table*}

We performed spectroscopy for the selected LBGs with Visible
Multi-Object Spectrograph (VIMOS; \citealt{lef03}) on Very Large
Telescope (VLT) under the program ID of 081.A-0081(A) (PI: A.~K.~Inoue). 
We used the LR-Blue/OS-Blue setting where the spectral resolving power 
$R=\lambda/\Delta\lambda\simeq180$ and the pixel scale is 5.3\,\AA/pix. 
With VIMOS, we can observe 4 quadrants with about 2\,arcmin
separation simultaneously in one pointing, 
for a total FoV of about $14\times16$ arcmin$^2$ in each pointing. 
We observed 2 pointings whose coordinates, observed month, exposure time, and seeing 
are listed in Table~1. We had 623 LBGs satisfying the selection criteria
described above in the two FoVs, out of which we observed 163 objects. 
We avoided galaxies which had spectroscopic redshifts previously
obtained. This would cause a bias in the galaxy selection, but we did not
correct it in the expected redshift distribution described in the next
subsection.

The data reduction was done with the VIMOS pipeline 
recipes\footnote{http://www.eso.org/sci/software/pipelines/vimos/}
and the NOAO/IRAF\footnote{http://iraf.noao.edu/}. We used the pipeline
software only to make bias and flat frames. Other steps were done with
IRAF through the standard manner (see \citealt{kou11} in detail). During
the data reduction, we found that the data quality in a quadrant of
FoV-1 was very low because there was no object in a half of the
images obtained from this quadrant. This might be caused by a
displacement of the slit mask for the quadrant in observations. We
decided to discard this quadrant unfortunately. 

To produce the one-dimensional spectrum of each LBG, we extracted
spatially 4 pixels tracing the object continuum from the background
subtracted and median coadded two-dimensional spectral image, and summed
them up. Given the spatial pixel scale of $0.205''$/pix, the extracted
spatial scale is 0.82\,arcsec which was chosen to maximize the
S/N ratio for the continuum rather than to collect the total flux of the objects.

We have determined Ly$\alpha$ emission/absorption and
metal absorption redshifts ($z_{\rm Ly\alpha}$ and $z_{\rm metal}$) 
of the LBGs by eye after applying a 5-pix box-car 
smoothing to the one-dimensional spectra. The 5-pix almost 
corresponds to the spectral resolution of the VIMOS LR-Blue setting with
$R=180$ and 5.3\,\AA/pix. The spectral features which we searched for were 
Ly$\alpha$ emission/absorption (1215.67\,\AA\ in the source rest-frame),
Ly$\beta$ (1025.72\,\AA) and Ly$\gamma$ (972.54\,\AA) absorption lines, 
Si {\sc ii} (1260.42\,\AA), O {\sc i} (1302.17\,\AA), Si {\sc ii} (1304.37
\AA), C {\sc ii} (1334.53\,\AA), Si {\sc iv} (1393.76 and 1402.77\,\AA),
Si {\sc ii} (1526.71\,\AA), C {\sc iv} (1548.20 and 1550.78\,\AA), Fe {\sc
ii}  (1608.45\,\AA), and Al {\sc ii} (1670.79\,\AA) absorption lines, and He
{\sc ii} (1640.4\,\AA) emission. We also searched [O {\sc ii}] (3727.5\,\AA), 
H$\beta$ (4861.3\,\AA), [O {\sc iii}] (4958.9 and 5006.8\,\AA), and
H$\alpha$ (6562.8\,\AA) emission lines as a signature of low redshift
contamination.

\begin{figure}
 \begin{center}
  \includegraphics[width=70mm]{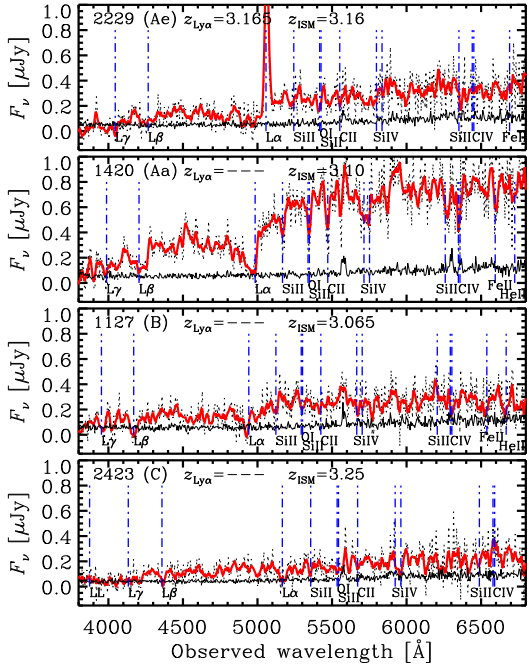}
 \end{center}
 \caption{Example one-dimensional spectra of the four redshift
 categories. The thick solid and thin dotted lines are the object
 spectra with and without a 5-pix box-car smoothing, respectively. 
 The thin solid lines are the root-mean-square spectra of the
 sky-subtracted background. The vertical dot-dashed lines show the
 wavelengths of some emission/absorption lines and the Lyman limit (LL). 
 (a) Ae: clear Ly$\alpha$ emission is identified, 
 (b) Aa: clear Ly$\alpha$ absorption and
 several clear metal absorption lines are identified, (c) B: Ly$\alpha$
 absorption and a few metal absorption lines are identified, and (d) C: 
 possible Ly$\alpha$ emission/absorption and/or possible metal absorption 
 are identified.}
 \label{example}
\end{figure}

We have classified the redshifts into four categories: Ae (clear Ly$\alpha$
emission is identified), Aa (clear Ly$\alpha$ absorption and several
clear metal absorption lines are identified), B (Ly$\alpha$ absorption and a
few metal absorption lines are identified), and C (possible Ly$\alpha$
emission/absorption and/or possible metal absorption are identified). 
Figure~\ref{example} shows example spectra of the four categories. 
The resultant numbers of $z\sim3$ LBGs are summarized in Table~2. 


\begin{table}
 \centering
 \begin{minipage}{70mm}
  \caption{Summary of VIMOS redshift survey results in the SSA22 field.}
  \begin{tabular}{lc}
   \hline
   Area (arcmin$^2$) & 322 \\
   $N_{\rm cand}$$^a$ & 623 \\
   $N_{\rm spec}$$^b$ & 163 \\
   $N_{\rm Ae}$$^c$ & 39 \\
   $N_{\rm Aa}$$^c$ & 18 \\
   $N_{\rm B}$$^c$ & 21 \\
   $N_{\rm C}$$^c$ & 21 \\
   \hline
  \end{tabular}\\
  $^a$ Number of the photometric LBG candidates.\\
  $^b$ Number of the objects observed in the spectroscopy.\\
  $^c$ Number of the objects classified into each category.
 \end{minipage}
\end{table}


According to \cite{ade05}, the redshift of the Ly$\alpha$ emission line
is slightly redshifted compared to those of the nebular
emission lines in the rest-frame optical which are more reliable as the
systemic redshifts. On the other hand, the redshifts of the metal
absorption lines are slightly blueshifted compared to 
those of the optical nebular lines. Then, we adopt the 
calibration formulae proposed by \cite{ade05} to estimate the systemic
redshifts of the LBGs. For LBGs with the Ly$\alpha$ emission line, we
adopt
\begin{equation}
 z_{\rm sys} = z_{\rm Ly\alpha} - 0.0033 - 0.0050 (z_{\rm Ly\alpha}-2.7)\,,
  \label{zsys1}
\end{equation}
and for LBGs without the Ly$\alpha$ emission line, we adopt
\begin{equation}
 z_{\rm sys} = z_{\rm metal} + 0.0022 + 0.0015 (z_{\rm metal}-2.7)\,.
  \label{zsys2}
\end{equation}
The uncertainties of $z_{\rm Ly\alpha}$ and 
$z_{\rm metal}$ are $\approx0.005$ estimated from the wavelength pixel
scale of 5.3~\AA/pix. The uncertainties of equations~(1) and (2) are 
$\approx0.003$ \citep{ade05}. If we take the summation in quadrature,
the uncertainty of $z_{\rm sys}$ is $\approx0.006$.  
Although there are updates of these formulae, for example, by
\cite{ste10} and \cite{tur14}, the accuracy of the redshifts in
equations (\ref{zsys1}) and (\ref{zsys2}) is sufficient for our analysis
because we make a composite in the observer's rest-frame not in the galaxies' rest-frame.

\subsection{Redshift distribution of the LBGs}


\begin{figure}
 \begin{center}
  \includegraphics[width=70mm]{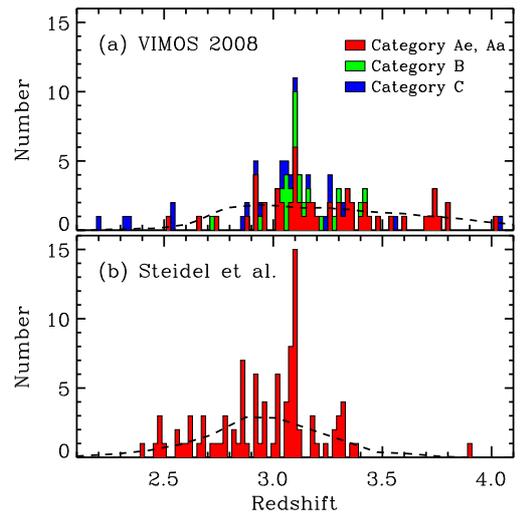}
 \end{center}
 \caption{Spectroscopic redshift distributions of LBGs by (a) the VIMOS
 survey in 2008 and (b) Steidel et al.~(2003). For the VIMOS survey, we
 show the different redshift categories (i.e., qualities) by different
 colors as indicated in the panel. The dashed lines are the expected
 distributions in each survey assuming a uniform distribution of
 galaxies.}
 \label{zdist}
\end{figure}

The redshift distribution of the LBGs in our survey is shown in the top
panel of Figure~\ref{zdist}, while that of the survey of \cite{ste03} in the
same field is shown in the bottom panel of the figure. We find several
redshift spikes in these distributions. In particular, 
the strongest peak is $z=3.1$. This is the redshift of the huge overdensity
of galaxies previously known in this field  
\citep{ste98,ste00,hay04,mat04,mat05,yam12,sae15,top18}.

In the top panel, we also show the expected number distribution from our
photometric selection criteria if the galaxies were distributed uniformly 
in the Universe. The expected number of galaxies at the redshift between
$z$ and $z+\Delta z$ is given by  
\begin{equation}
 N_{\rm exp}(z) = \int_{M_{\rm min}}^{M_{\rm max}} \phi(M) 
  \int_z^{z+\Delta z} C(z',M) \frac{dV}{dz}(z')~dz'~dM\,,
  \label{ngal}
\end{equation}
where $\phi(M)$ is the number density of LBGs with the absolute
magnitude $M$ (i.e., luminosity function) , $C(z,M)$ is the selection
efficiency for an object with $M$ at $z$ (i.e., completeness), and
$dV/dz(z)$ is the volume element. We assume the luminosity function of
$z\sim3$ LBGs reported by \cite{ste99} for $\phi(M)$. The
integration boundaries are set to be $M_{\rm min}=-24.0$  and 
$M_{\rm max}=-17.0$ which do not affect the result very much because 
the magnitude limit described in section 2.1 is included in $C(z,M)$. 

To obtain $C(z,M)$, we have 
performed a Monte-Carlo simulation of our color selection. First, we
generated a large number of mock galaxies having $z$, $M$, and a
spectrum. For the spectrum, we prepared 4 types depending 
on the Ly$\alpha$ strength as reported by \cite{ste03}. These four
spectra were produced based on the population synthesis model GALAXEV 
\citep{bru03} to extend shorter and longer wavelengths than those
observed by \cite{ste03}. We then applied a mean IGM transmission by
\cite{ino05}. Second, we calculated apparent magnitudes from $u^*$ to
$z'$ of the mock galaxies based on their spectra, redshift $z$, and
$M$. Then, we randomly added Gaussian errors based on the limiting
magnitude in each band to the apparent magnitudes. In this step, we
mixed the 4 types of spectra with an equal weight. Finally, we applied
the color (and magnitude) selection to them and counted the number
fraction selected as a function of ($z$, $M$).

Since we did not observe all galaxies satisfying the color selection
and could measure redshifts for only a part of the observed galaxies, we
cannot compare the expected number of galaxies in equation~(\ref{ngal}) with the
obtained redshift distribution directly. Therefore, we normalized the
expected number distribution by the total number of galaxies for which we
measured their redshifts successfully, following \cite{ste98,ste00,ste03}. 

The expected distribution in Steidel et al.'s survey in the bottom panel
of Figure~\ref{zdist} is empirically obtained from the sum of 17
different fields of their spectroscopic survey. The distribution is
again normalized by the total number of the redshifts in the SSA22
field. The shape of their empirical distribution function of LBGs is
quite different from our expected one. This is because they used a
different filter set and applied a different color selection from those we did.

\section{Observer's frame composite spectrum}

Using the VIMOS LBG spectra, we examine the IGM H~{\sc i}
fluctuation along the sight-line of the SSA22 field. The
continuum S/N ratio per wavelength element in the
so-called ``DA'' (Depression at Ly$\alpha$) wavelength range 
of the individual LBG spectra distributes 
from 1 to 8 and the median is about 3. Thus, we adopt 
a stacking technique to increase the S/N ratio. Since we are 
investigating the IGM, we make an observer's frame
composite spectrum. This composite analysis also means that we will
examine an average of the H~{\sc i} fluctuation over the observing
field. In this analysis, we restrict ourselves to the spectra
categorized as Ae, Aa, and B to avoid possible contamination of
lower-$z$ spectra in the category C. In addition, we remove one object
in the category Ae (Slit \#2408) because of its low S/N in the continuum 
(S/N $\sim0.2$ in the DA range). Therefore, we use 77 LBG spectra in total.

Here, we propose a new method to make an observer's frame composite rather
than a simple sum of the spectra over the whole wavelength coverage as
done by \cite{cuc14} in order to avoid contamination of galaxies'
Ly$\alpha$ emission/absorption line and many interstellar absorption
lines in the resultant composite spectrum. The intergalactic Ly$\beta$
absorption lines also contaminate the spectrum for the highest redshift
LBGs. In order to isolate the intergalactic H~{\sc i} Ly$\alpha$ 
absorption, we stack only the wavelength range between Ly$\alpha$ and
Ly$\beta$ in the source rest-frame, the so-called DA range. 
There is a small contamination of narrow absorption lines 
by other atoms like C, O, and so on in the IGM at lower redshift. 
We neglect it because we can not identify these lines in our low-resolution 
and low S/N ratio spectra. However, this effect has been estimated to be 
as small as 3\% in terms of the transmission at $z=3$ \citep{fau08}. 
On the other hand, some metal absorption lines in the stellar photosphere 
and in the ISM of galaxies may contaminate in the spectral range. 
In the rest-frame composite spectra of $z\sim2$--3 galaxies reported 
by \cite{sha03} and \cite{ste10}, 
we can identify S {\sc iv} $\lambda$1063, N {\sc ii} $\lambda$1084, and 
C {\sc iii} $\lambda$1178 lines between Ly$\alpha$ and Ly$\beta$. Then,
we define the DA range  as 1070--1170\,\AA\ with a narrow mask of 6.5
\AA\ ($=5$\,pix at $z\sim3$) around 1084\,\AA\ in the source rest-frame to
avoid these metal lines as well as any effects of broad Ly$\alpha$ and
Ly$\beta$ absorption features of the emitting galaxy itself. The final wavelength width
to be used in the composite is about 370\,\AA\ in the observer's frame and
about 70 pixels, corresponding to $\Delta z\approx0.3$ for H~{\sc i}
Ly$\alpha$.

\subsection{Procedure to make the DA range composite spectrum}

The procedure for making the observer's frame composite spectrum consists
of the following 3 steps:

\begin{enumerate}
 \item Clip out the DA range in the source rest-frame from each
       one-dimensional observed spectrum 
       without any smoothing: $f^{\rm obs}_{\nu_{\rm DA}}$.

 \item Make a linear fit of the clipped-out spectrum\footnote{We used
       all the wavelength pixels in the DA range, except for several
       pixels possibly affected by N {\sc ii} $\lambda$1084 absorption,
       for this linear fit because the possible IGM H~{\sc i} absorption
       enhancement would be narrow enough relative to the entire DA
       range and would not affect the fit very much. Indeed, we have
       confirmed that the $z=3.1$ H~{\sc i} enhancement is robust 
       even if we applied $N\sigma$-clipping to the linear fitting. 
       For $N=0.5$, 1, 2, or 3, we have found $-35$\%, $+16$\%, $-7$\%, 
       or $-3$\% change in the {\it excess} equivalent width,
       respectively, which are comparable or smaller than the
       uncertainty obtained by a bootstrap method.}
       and normalize it by the fit: 
       $\tilde{f}_{\nu_{\rm DA}}
       =f^{\rm obs}_{\nu_{\rm DA}}/f^{\rm fit}_{\nu_{\rm DA}}$. 

 \item Make a median or 3-$\sigma$ clipping average composite of the 
   normalized spectra in the observer's frame. 

\end{enumerate}

After making the composites, we apply a 5-pix boxcar smoothing to the
spectra. The choice of 5 pixels is based on the spectral resolution
of our VIMOS observations with LR-Blue/OS-Blue ($R\simeq180$ and pixel
scale of 5.3\,\AA/pix). The 5 pixels also correspond to a scale of 
about 20 comoving Mpc at $z=3.1$ for H~{\sc i} Ly$\alpha$. This scale is
very similar to the transverse scale of the filaments in the overdensity
structure traced by LAEs at $z=3.1$ \citep{hay04,yam12}. Therefore, we
can examine a structure larger than this scale along the sight-line in the
composite spectrum.

The physical meaning of the normalized spectrum, $\tilde{f}_\nu$, is the
fluctuation of the H~{\sc i} absorption relative to an average one as we
find from the following discussion. If we express the IGM H~{\sc i}
Ly$\alpha$ optical depth along a sight-line as
\begin{equation}
 \tau_{\nu(z)}^{\rm IGM} = \langle \tau_{\nu(z)}^{\rm IGM} \rangle 
  + \delta \tau_{\nu(z)}^{\rm IGM}\,,
\end{equation}
where $\langle \tau_{\nu(z)}^{\rm IGM} \rangle$ is the cosmic mean
optical depth at the redshift $z$ and $\delta \tau_{\nu(z)}^{\rm IGM}$ 
is the fluctuation relative to the mean, the observed flux becomes 
\begin{equation}
 f_\nu^{\rm obs} = e^{-\langle \tau_{\nu(z)}^{\rm IGM} \rangle}
  e^{-\delta \tau_{\nu(z)}^{\rm IGM}} f_\nu^{\rm cont}
  =\langle T_\nu^{\rm IGM} \rangle e^{-\delta \tau_{\nu(z)}^{\rm IGM}} 
  f_\nu^{\rm cont}\,,
\end{equation}
where $f_\nu^{\rm cont}$ is the continuum before the IGM absorption, and
$\langle T_\nu^{\rm IGM} \rangle$ is the cosmic mean IGM transmission. If
the period of the fluctuation $\delta \tau_{\nu(z)}^{\rm IGM}$ is short
enough relative to the DA range, the linear fit spectrum can be
expressed as 
\begin{equation}
 f_\nu^{\rm fit} = \langle T_\nu^{\rm IGM} \rangle f_\nu^{\rm cont}\,,
\end{equation}
because the variation by $\delta \tau_{\nu(z)}^{\rm IGM}$ is smoothed
out. Therefore, we obtain 
\begin{equation}
 \tilde{f}_\nu = e^{-\delta \tau_{\nu(z)}^{\rm IGM}}\,, 
\end{equation}
and 
\begin{equation}
 \delta \tau_{\nu(z)}^{\rm IGM} = -\ln \tilde{f}_\nu
\end{equation}

\subsection{Composite of sky-subtracted background}

The uncertainty caused by the fluctuation of the sky-subtracted
background can be measured by making a composite of the sky-subtracted
background spectra. We stack only the ``DA range'' of the sky-subtracted 
background based on each LBG's redshift, as follows. First, in each
sky-subtracted two-dimensional spectrum, we define 10 pixels along the 
spatial direction as the background region, avoiding
pixels possibly including object flux. 
Then, we calculate the sum of the background brightness, $I^{\rm back}_{\nu_{\rm DA}}$,  
of the 10 pixels for each wavelength element and scale it to the $N_{\rm pix}=4$ 
extraction so as to be equivalent to the object spectra. 
Namely, the sky-subtracted background spectrum 
$f^{\rm back}_\nu = (N_{\rm pix}/10)^{1/2} \sum_{i=1}^{10} I^{\rm back}_\nu$.
Next, we stack $f^{\rm back}_\nu$ in the almost same way as the object spectra described 
in the previous subsection. However, $f^{\rm back}_\nu$ distributes around zero 
because it is the residual of the sky subtraction. The linear fit and normalization 
in the step (ii) of the composite procedure causes erroneously large fluctuation. 
Thus, we replace the clipped-out spectrum of the step (i) with 
\begin{equation}
 f^{\rm obs'}_{\nu_{\rm DA}} = f^{\rm fit}_{\nu_{\rm DA}} + f^{\rm back}_{\nu_{\rm DA}} \,.
\end{equation}
The fit spectrum $f^{\rm fit}_{\nu_{\rm DA}}$ is a linear fit of each object spectrum
within the DA range: the same one used as the normalization in the step
(ii) for the object composite. In the step (ii) of this background
composite, the replaced clipped-out spectrum is normalized by another
linear fit function obtained from the replaced spectrum as:
\begin{equation}
 \tilde{f}_{\nu_{\rm DA}} = 
  f^{\rm obs'}_{\nu_{\rm DA}} / f^{\rm fit'}_{\nu_{\rm DA}}\,.
\end{equation}
The last step in the procedure is the same. We also apply a 5-pix boxcar
smoothing to the resultant composite.

\subsection{Monte-Carlo simulation of the composite}

To estimate the fluctuation of the composite spectrum caused by the Ly$\alpha$ forest (LAF) 
and observational errors (i.e., background fluctuation), we perform a
Monte-Carlo simulation which generates a large number of mock observed spectra: 
\begin{equation}
 f_\nu^{\rm MC} = T_\nu^{\rm IGM} f_\nu^{\rm cont} 
  + \delta_\nu^{\rm back}\,,
\end{equation}
where $\nu$ is the frequency in the observer's frame, 
$T_\nu^{\rm IGM}$ is the intergalactic transmission, $f_\nu^{\rm cont}$
is the continuum spectrum of galaxies, and $\delta_\nu^{\rm back}$ is
the observational error caused by the sky-subtracted background
fluctuation.

The intergalactic transmission $T_\nu^{\rm IGM}$ is generated by the
Monte-Carlo simulation model developed by \cite{ino08}. We adopt the
latest version in which the statistics of IGM absorbers is updated
\citep{ino14} although this update does not affect the transmission in
the DA range significantly. In the simulation, we mimic our VIMOS
observations as follows. The wavelength pixel scale of the observations
is 5.3\,\AA/pix. Since this resolution is too coarse to resolve fine
absorption lines by the LAF, we adopt ten times finer resolution in the
calculations: 0.53\,\AA\ which corresponds to $\approx0.1$\,\AA\ in the
rest-frame at $z=3$. This resolution is 
fine enough to give a $\sim1\%$ accuracy in transmission compared to
that calculated with a 0.01 \AA\ resolution \citep{ino08}. 
Then, we apply a smoothing with a
Gaussian function whose FWHM is $\Delta\lambda=\lambda/R$, where the
resolving power of $R=180$ and a typical wavelength of $\lambda=5000$
\AA\ in our observations. Finally, we average 
the resulting transmission values for intervals of 10
pixels to match the pixel scale of the VIMOS observations. 

We note here that there is neither sight-line (i.e., redshift) nor spatial
correlation of absorbers in our Monte-Carlo simulation. The absorbers
follow their empirical distribution function but are completely randomly
located from each other. However, the absorbers do correlate with
themselves in the real Universe
\citep[e.g.,][]{zuo94,cri95,mei95,cro99,mcd00}. Thus, our simulation 
underestimates the fluctuation of the IGM. On the other hand, we still
have the sensitivity to detect the real IGM fluctuation with this random IGM
simulation, examining if the resultant composite spectrum has
fluctuation significantly larger than that expected from random.

\begin{figure}
 \begin{center}
  \includegraphics[width=70mm]{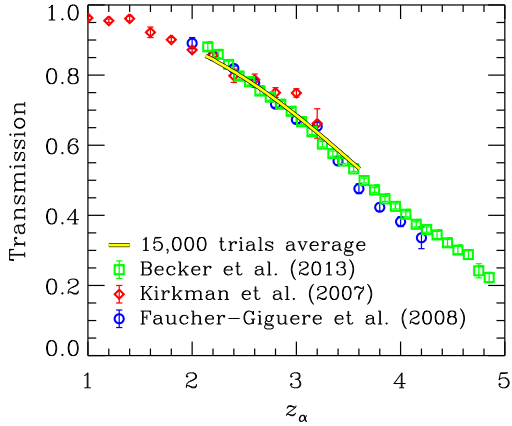}
 \end{center}
 \caption{Intergalactic transmission as a function of the
 absorber's redshift $z_\alpha=\lambda_{\rm obs}/\lambda_\alpha-1$,
 where $\lambda_{\rm obs}$ is the observed wavelength and 
 $\lambda_\alpha=1215.67$\,\AA\ is the H~{\sc i} Ly$\alpha$ wavelength. 
 The data points with error bars are taken from the literature as shown
 in the panel. The solid curve is the mean transmission obtained
 from our Monte-Carlo simulation described in equation~(14).}
 \label{La}
\end{figure}

The continuum $f_\nu^{\rm cont}$ in the DA range is never observed
because it is modulated by the LAF. Here we simply assume 
$f_\nu^{\rm cont}$ to be constant in the DA range: 

\begin{equation}
 f_{\nu_{\rm DA}}^{\rm cont} = 
  \frac{\int_{\Delta\nu_{\rm DA}} f_\nu^{\rm obs}
  /\langle T_\nu^{\rm IGM} \rangle d\nu}
  {\Delta\nu_{\rm DA}} \,,
\end{equation}
where $f_\nu^{\rm obs}$ is the observed spectrum, 
$\langle T_\nu^{\rm IGM} \rangle$ is a mean intergalactic
transmission, and $\Delta\nu_{\rm DA}$ is the frequency
interval of the DA range. The mean transmission is assumed to be 
\begin{equation}
 \langle T_\nu^{\rm IGM} \rangle = 
 e^{-\langle \tau_\nu^{\rm IGM} \rangle}\,,
\end{equation}
and 
\begin{equation}
 \langle \tau_\nu^{\rm IGM} \rangle = 0.427
 \left(\frac{\lambda}{5000~{\rm \AA}}\right)^{3.7}\,,
\end{equation}
where the observed wavelength $\lambda=c/\nu$ with the
speed of light $c$. This mean intergalactic optical depth
is obtained by averaging the results from the Monte-Carlo
simulation and matches the observed optical depths as
shown in Figure~\ref{La}.


The background fluctuation $\delta_\nu^{\rm back}$ is obtained from the
sky-subtracted two-dimensional spectra. As described in the previous
subsection, we identify spatially 10 pixels as the background region in
each two-dimensional spectrum. Then, we calculate the 
root-mean-square (r.m.s.) spectrum of the sky-subtracted background 
from the 10 pixels, which we denote $e^{\rm back}_\nu$. A typical value
of $e^{\rm back}_\nu$ is $\approx28$ nJy per one spatial pixel at 5000
\AA. Then, we draw $\delta_\nu^{\rm back}$ randomly from a Gaussian
distribution with the mean of 0 and the standard deviation of 
$\sqrt{N_{\rm pix}}\times e^{\rm back}_\nu$, where $N_{\rm pix}=4$ is
the number of the spatial pixels in the object spectrum extraction.

\begin{figure}
 \begin{center}
  \includegraphics[width=70mm]{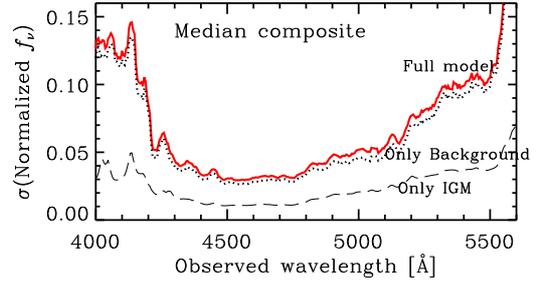}
 \end{center}
 \caption{The standard deviation spectrum of the 15,000 observer's frame 
 median composite spectra generated by the Monte-Carlo simulation (solid line). 
 Breakdown of the contributions of the sky-subtracted background 
 (dotted line) and IGM fluctuations (dashed line) to the standard deviation.}
 \label{MCfluc}
\end{figure}

We have generated 15,000 mock observed spectra for each LBG. 
These mock spectra are  
processed by the same procedure as the real observed spectra described
in section~3.1, and then, we obtain composite spectra. The distribution of the
composite (normalized) flux densities in each wavelength element is well
described by a Gaussian function around unity. Note that the spectra
were normalized in the composite procedure, and then, the mean should be
about unity by construction. The standard deviation of the distribution gives us an
estimate of the uncertainty at each wavelength in the resultant composite 
spectrum. Figure~\ref{MCfluc} shows the breakdown of the contributions
of the intergalactic absorbers and the background fluctuation to the
standard deviation. We find that the background contribution is dominant. 

In addition, we tried two other methods for estimating $f_\nu^{\rm cont}$ of 
equation (11): a linear continuum in the DA range and a power-law fit at 
longer wavelengths and an extrapolation to the DA range. 
The resultant standard deviation spectra were very similar to that of the constant 
case shown in Figure\,\ref{MCfluc}. In the following, we adopt the simplest 
constant DA case.


\section{Result}


\begin{figure*}
 \begin{center}
  \includegraphics[width=140mm]{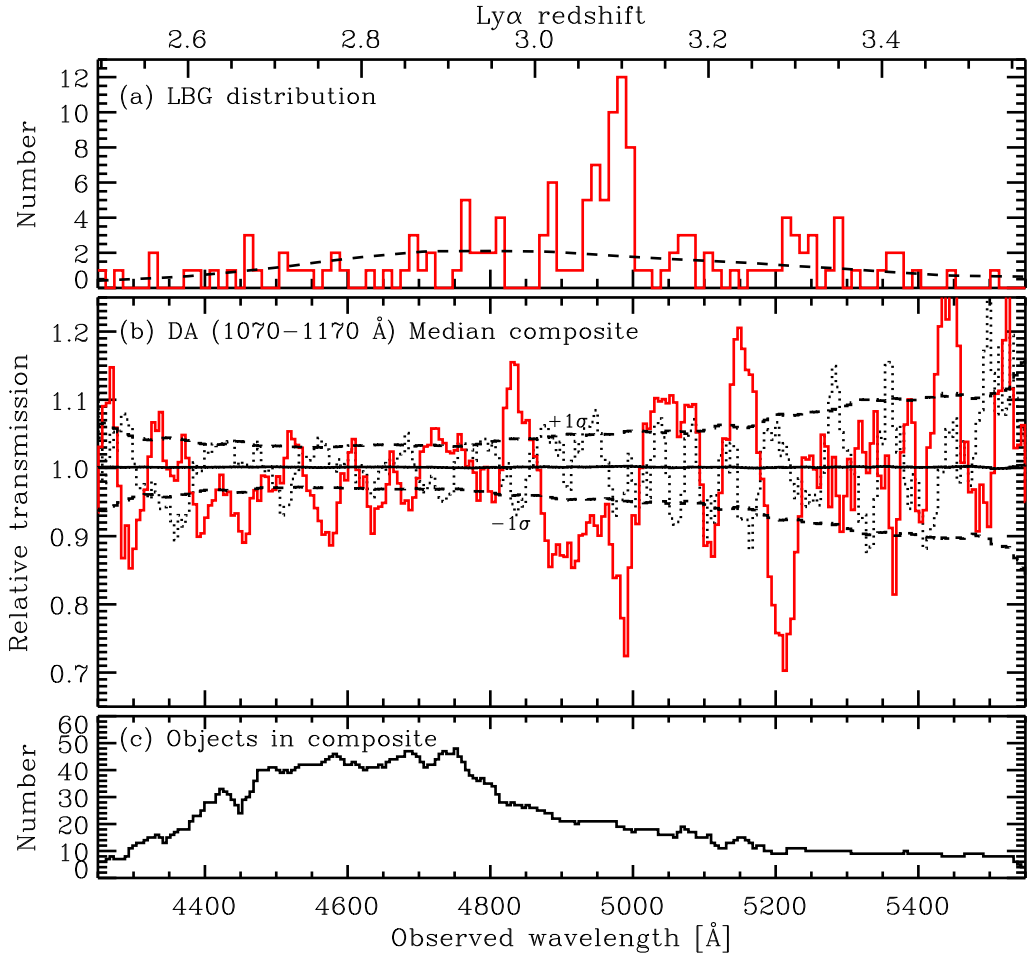}
 \end{center}
 \caption{Observer's frame composite spectrum of LBGs in the SSA22 
 field. (a) Spectroscopic redshift distribution of the
 LBGs whose redshifts are categorized as Ae, Aa and B from our VI08  
 observation and the LBGs from Steidel et al.~(2003). 
 The dashed line is the distribution expected when the galaxies
 distribute uniformly. 
 (b) The IGM transmission relative to the mean. The dotted line is the 
 composite of the sky-subtracted background. The almost horizontal solid line with 
 the intercept at 1.0 and the dashed lines indicate the mean and $\pm1$\,$\sigma$ 
 for one pixel from a Monte-Carlo simulation of the composite procedure, respectively. 
 (c) The number of objects used for the composite at each wavelength element.} 
 \label{DAcomp} 
\end{figure*} 

We show in Figure~\ref{DAcomp} the resultant observer's frame median
composite spectrum. We also obtain a similar result from the 3-sigma clipping 
average composite. The displayed range is the wavelength where the
number of LBGs used in the composite is larger than 7 as shown in
the bottom panel. In the middle panel, the composite spectra are shown
by the solid line, while the sky-subtracted background composite is
shown by the dotted line. We can see larger fluctuations in the
object composite than in the background composite, indicating 
the reality of these features in the object composite. The almost
horizontal line around unity actually shows the mean of the 15,000
composites generated by the Monte-Carlo simulation. The standard
deviation in each wavelength pixel of these simulated
composites is shown by the dashed lines which are well matched with the
fluctuation of the background composite. This is consistent with
what we have seen in Figure~\ref{MCfluc}; the background fluctuation
dominates the fluctuation by random IGM absorbers. 

The top panel shows the redshift distribution of the LBGs reported by
\cite{ste03} and our VI08 survey. 
The dashed line shows the expected number 
for a universe where LBGs were distributed uniformly, 
which is estimated from a number-weighted average of 
the selection functions of VI08 and \cite{ste03} shown in Figure~\ref{zdist}. 
We can find some spikes and voids in the distribution 
and the most prominent spike is the known PC at $z=3.1$
\citep{ste98,ste00,hay04,mat04,mat05,yam12}.  

Very interestingly, we can see a strong dip in the object composite at
the wavelength exactly corresponding to the PC H~{\sc i}
Ly$\alpha$. Furthermore, some peaks and dips in the object composite 
seem to correlate with voids and spikes in the LBG redshift distribution,
respectively, especially at wavelengths longer than 4,800\,\AA\ or
redshift $z>2.95$. As described in equation (7), the observer's frame
composite obtained in this paper is equivalent to the IGM transmission
spectrum normalized by its mean. Thus, the dip/peak of the object
composite corresponds to more/less absorption in the IGM than the
mean at that redshift.



In this paper, we focus on two sharp absorption dips which have the minimum 
transmission less than 0.8, together with two sharp
transparency peaks greater than 1.15 in Figure~\ref{DAcomp}. 
All dips and peaks selected are also required to have 10 or more
sight-lines. Their redshifts are 3.10 and 3.28 for the dips 
as well as 2.98 and 3.24 for the peaks respectively. 
There is another interesting absorption dip around at 
$z=3.04$, probably corresponding to the LBG density peak at the same
redshift. As \cite{top18} discussed, this LBG overdensity is another
PC. However, the absorption dip is shallower but wider than those
selected above. Therefore, we will defer to examine this feature until
more data are available for this.

\subsection{EWs and significance of the peaks and dips in the transmission spectrum}


There is a significant fluctuation in the observer's frame 
composite. This is equivalent to the fluctuation relative to the mean
IGM transmission because we have normalized individual LBG spectra 
during the composite procedure. The intrinsic galaxy spectrum in the DA
range which is used in the composites is smooth enough and can not produce 
such a fluctuation. To quantify the significance of the peaks and dips
against the mean IGM transmission, we define the {\it excess} equivalent
width (EW) as 
\begin{equation}
 EW_{\rm exc} \equiv \sum_{\lambda_1 \leq \lambda \leq \lambda_2} 
  (\tilde{f_\lambda}-1) \Delta \lambda \,,
\end{equation}
where $\tilde{f_\lambda}$ is the normalized composite flux 
density\footnote{$\tilde{f_\lambda}=f_\lambda/f^{\rm cont}_\lambda
=f_\nu/f^{\rm cont}_\nu=\tilde{f_\nu}$ which is the obtained composite
spectrum and the IGM transmission fluctuation divided by the mean
transmission as described in equation~(7), }, $\Delta \lambda=5.3$\,\AA\
is the width of the wavelength element, and $\lambda_1$ and $\lambda_2$
are the lower and upper wavelength boundaries to be summed up as a
feature, respectively. Then, let us define two different uncertainties
for $EW_{\rm exc}$ based on the standard deviation spectrum obtained by
the Monte-Carlo simulation, $\sigma_{\rm MC,\lambda}$. 
Note that $\sigma_{\rm MC,\lambda}$ is non-dimensional. 
One is the case without any correlation in the wavelength space: 
\begin{equation}
 \sigma^{EW}_{\rm noncor} = \sqrt{\sum_{\lambda_1 \leq \lambda \leq \lambda_2} 
  {\sigma_{\rm MC,\lambda}}^2 {\Delta \lambda}^2}\,.
\end{equation}
The other is the case with a complete correlation in the wavelength
space: 
\begin{equation}
 \sigma^{EW}_{\rm cor} = \sum_{\lambda_1 \leq \lambda \leq \lambda_2} 
  \sigma_{\rm MC,\lambda} \Delta \lambda \,.
\end{equation}

In the Monte-Carlo simulation, we have assumed neither correlation of
IGM absorbers nor correlation of the background fluctuation along 
wavelength (or redshift). However, there should be a redshift
correlation of IGM absorbers \citep[e.g.,][]{zuo94,cri95,mei95,cro99,mcd00}. 
Thus, $\sigma_{\rm MC,\lambda}$ tends to underestimate the fluctuation
in the real Universe. In this sense, the former, non-correlated
uncertainty, $\sigma^{EW}_{\rm noncor}$ would result in an underestimation.
On the other hand, the latter, completely correlated uncertainty,
$\sigma^{EW}_{\rm cor}$ would result in considerable overestimation when 
the background noise is random and dominates $\sigma_{\rm MC,\lambda}$. 
We do not know how much the absorbers' correlation enhances their contribution in 
$\sigma_{\rm MC,\lambda}$ quantitatively, at the moment, while the
contribution is minor in the no absorbers' correlation case, as seen in
Figure~\ref{MCfluc}. 
In summary, the real uncertainty should be bracketed by these extreme cases.
Then, we define the two S/N ratios for $EW_{\rm exc}$ 
\begin{equation}
 (S/N)_{\rm max} = \frac{|EW_{\rm exc}|}{\sigma^{EW}_{\rm noncor}}\,,
  \label{maxsn}
\end{equation}
and
\begin{equation}
 (S/N)_{\rm min} = \frac{|EW_{\rm exc}|}{\sigma^{EW}_{\rm cor}}\,,
  \label{minsn}
\end{equation}
as the {\it maximum} and {\it minimum} S/N ratios, respectively.


\begin{table*}
 \centering
 \begin{minipage}{140mm}
  \caption{A summary of significance of the peaks and dips in the transmission spectrum.}
  \begin{tabular}{lcccc}
   \hline
   Redshift & \multicolumn{2}{c}{Median} 
   & \multicolumn{2}{c}{Average} \\
   peak/dip & $(S/N)_{\rm min}$$^a$ & $(S/N)_{\rm max}$$^a$ 
	& $(S/N)_{\rm min}$$^a$ & $(S/N)_{\rm max}$$^a$ \\
   \hline
   2.98 peak &  2.86 (5pix) & 6.64 (7pix)  
 	     &  4.00 (5pix) & 9.31 (9pix) \\
   3.10 dip & 3.84 (5pix) & 8.98 (11pix) 
	    & 4.39 (5pix) & 9.81 (5pix) \\
   3.24 peak & 2.91 (5pix) & 6.91 (7pix) 
	     & 2.73 (5pix) & 6.12 (7pix) \\
   3.29 dip  & 3.26 (5pix) & 7.98 (7pix) 
	     & 1.81 (5pix) & 4.56 (9pix) \\
   \hline
  \end{tabular}\\
  $^a$ See the definitions of equations (18) and (19). The number in the
  parenthesis is the integrated pixels used in the S/N calculations. 

 \end{minipage}
 \label{peakdipdet}
\end{table*}

\begin{table*}
 \centering
 \begin{minipage}{165mm}
  \caption{Properties of the peaks and dips in the transmission spectrum.} 
  \begin{tabular}{lccccccccl}
   \hline
   Redshift & Wavelength & \multicolumn{2}{c}{Fiducial range} 
   & Number & Number & Number & \multicolumn{2}{c}{$EW_{\rm exc}$ [\AA]$^b$} & \\
   peak/dip & [\AA] & $\lambda_1$ [\AA] & $\lambda_2$ [\AA] 
	& of pixels & of sight-lines & of LBGs$^a$ & (Median) & (Average) & Remarks \\
   \hline
   2.98 peak & 4836.5 & 4820.6 & 4863.0 & 9 & 30 & 0 (8.2)
    & ${\phantom -}4.3\pm2.1$ & ${\phantom -}7.0\pm2.0$ & Void \\
   3.10 dip & 4984.9 & 4963.7 & 5006.1 & 9 & 21 & 34 (7.1)
    & $-7.0\pm2.3$ & $-7.1\pm2.0$ & Proto-cluster \\
   3.24 peak & 5154.5 & 5128.0 & 5175.7 & 10 & 17 & 2 (6.3)
    & ${\phantom -}6.3\pm3.5$ & ${\phantom -}5.3\pm2.6$ & Void? \\
   3.29 dip & 5207.5 & 5181.0 & 5234.0 & 11 & 12 & 12 (6.2)
    & $-7.3\pm7.3$ & $-5.5\pm4.9$ & Overdensity? \\
   \hline
  \end{tabular}\\
  $^a$ The number in the parenthesis is an expectation from a random
  distribution of galaxies.\\
  $^b$ {\it Excess} equivalent width in the observer's frame defined by
  equation (15) measured by a bootstrap method.
 \end{minipage}
 \label{peakdipEW}
\end{table*}

For the four dips and peaks in the transmission spectrum selected above, 
the significance by equation~(18) and (19) are summarized in Table~3 
for each case of median and (3-$\sigma$ clipping) average composites.
We have measured their {\it excess} EWs defined by equation~(15)  
by adopting a bootstrap method \citep[e.g.,][]{pre92};
before the step (iii) in the composite procedure described in section~3.1, we
insert one step of random resampling of the normalized DA spectra with
duplication. Then, we repeat the process 10,000 times. In this 
estimation, the wavelength range of the each dip/peak (from $\lambda_1$
to $\lambda_2$) is fixed to the range determined to cover the whole
structure of the dip/peak in the observed composite shown in
Figure~\ref{DAcomp}. The resultant {\it excess} EWs, 1\,$\sigma$ errors 
and the wavelength ranges are summarized for the four dips and peaks 
in Table~4. 
In the next subsection, we investigate their dips and peaks in detail. 


\subsection{Prominent peaks and dips in the transmission spectrum}


\begin{figure*}
 \begin{center}
  \includegraphics[width=120mm]{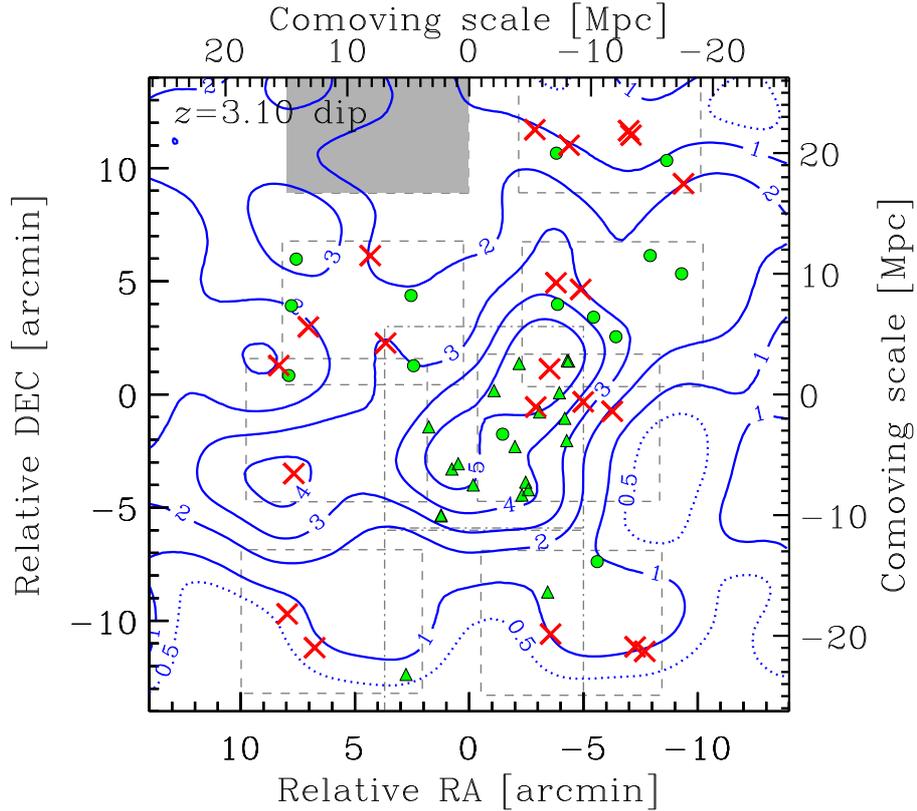}
 \end{center}
 \caption{Spatial distribution of the sight-lines probing the IGM
 between $z=3.07$ and $z=3.12$ (cross marks). The circles and 
 triangles are the LBGs whose redshifts are measured in the VI08 and 
 Steidel et al.~(2003) surveys, respectively, and are within the
 redshift range. The contour shows the surface number density map of
 LAEs at $z=3.06$--3.13 by Yamada et al.~(2012) and Hayashino et
 al.~(2004). The numbers along the contours indicate the
 density enhancement factor relative to the mean
 surface number density of LAEs in general fields at $z=3.1$: 
 $n_{\rm LAE}/\langle n_{\rm LAE} \rangle$. 
 The gray dashed and dot-dashed lines show the fields-of-view of VI08
 and Steidel~et~al.~(2003) observations, respectively. The shaded  
 north-east part is the unavailable quadrant of VI08 (see section~2.2).}
 \label{z3p10dip}
\end{figure*}


\begin{figure}
 \begin{center}
  \includegraphics[width=70mm]{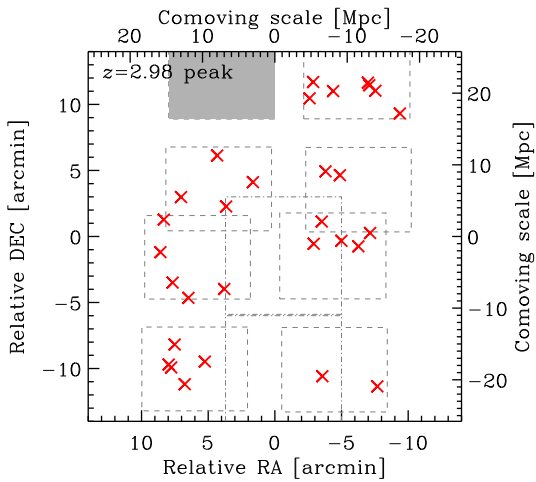}
 \end{center}
 \caption{The spatial distribution of the sight-lines  
 between $z=2.96$ and $z=3.00$. The gray lines and 
 shaded area are the same as in Figure~\ref{z3p10dip}.}
 \label{z2p98peak}
\end{figure}


\begin{figure}
 \begin{center}
  \includegraphics[width=70mm]{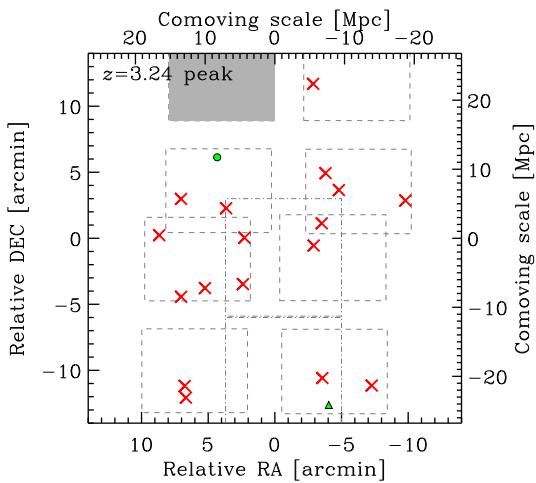}
 \end{center}
 \caption{Same as Figure~\ref{z2p98peak} but for the sight-lines 
 between $z=3.22$ and $z=3.26$. The LBGs in the redshift
 range are shown by the same symbols as in Figure~\ref{z3p10dip}. The
 FoVs are also shown as in Figure~\ref{z3p10dip}.}
 \label{z3p24peak}
\end{figure}


\begin{figure}
 \begin{center}
  \includegraphics[width=70mm]{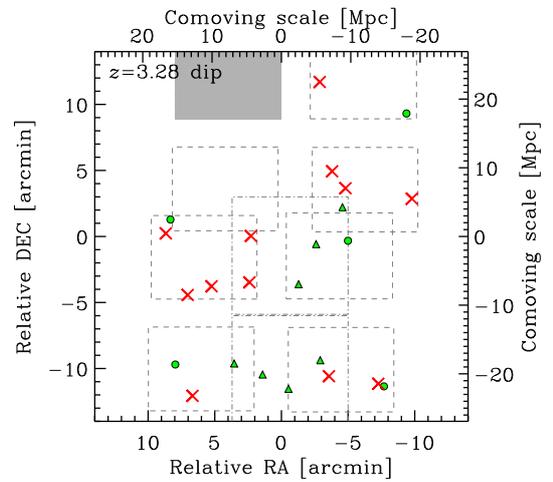}
 \end{center}
 \caption{Same as Figure~\ref{z3p24peak} but for the sight-lines 
 between $z=3.26$ and $z=3.31$.}
 \label{z3p28dip}
\end{figure}

\subsubsection{$z=3.10$ absorption dip}

\begin{table*}
 \centering
 \begin{minipage}{140mm}
  \caption{Dependence of $z=3.10$ IGM absorption dip on LAE
  overdensity and angular distance to the nearest {\it spectroscopic}
  LBG.}
  \begin{tabular}{lccccc}
   \hline
   & & & $\langle d_{1,\rm LBG} \rangle$$^c$ &
   \multicolumn{2}{c}{$EW^{\rm rest}_{\rm exc}$ [\AA]$^d$} \\ 
   Subsample & $N_{\rm SL}$$^a$ & $\langle \delta_{\rm LAE}
	   \rangle$$^b$ & [arcmin] & (Median) & (Average) \\
   \hline
   All & 21 & 1.19 & 1.98 & $-1.7\pm0.6$ & $-1.7\pm0.5$ \\
   \hline
   $\delta_{\rm LAE}\geq1$ & 7 & 3.14 & 1.13 & $-2.1\pm1.4$ & $-2.3\pm1.1$ \\
   $0.03\leq\delta_{\rm LAE}<1$ & 7 & 0.56 & 1.80 & $-2.2\pm1.1$ & $-2.3\pm1.0$ \\
   $\delta_{\rm LAE}<0.03$ & 7 & $-0.12$ & 3.00 & $-0.9\pm0.8$ & $-1.2\pm0.8$ \\
   \hline
  \end{tabular}\\
  $^a$ Number of sight-lines.\\
  $^b$ Average of the LAE overdensities at the positions of the sight-line, where 
  $\delta_{\rm LAE}=n_{\rm LAE}/\langle n_{\rm LAE} \rangle -1$. 
  \\
  $^c$ Average angular distance to the nearest {\it spectroscopic} 
  LBG from each sight-line.\\
  $^d$ Rest-frame {\it excess} equivalent width defined by equation~(15)
  measured by a bootstrap method.\\ 
 \end{minipage}
 \label{z3p10tab_d1}
\end{table*}

This absorption dip found at exactly the same redshift as the 
PC in the observing field is the most significant one 
detected ($\ga4$\,$\sigma$). In Figure~\ref{z3p10dip}, we show the spatial
distribution of the sight-lines contributing to the dip feature in the
composite as the cross marks. We also plot the positions of LBGs within
the redshift range corresponding to the dip feature ($3.07<z<3.12$)
taken from the catalogs by our VIMOS survey and \cite{ste03},
and the surface number density contour of LAEs at $z=3.06$--3.13 by
\cite{yam12}. These LAEs are selected with a narrowband filter in 
\cite{hay04} and its redshift coverage exceeds the lower redshift 
boundary of the dip feature, but most of the spectroscopic redshifts of 
the LAEs are around $z=3.09$ \citep{mat04}. The positions of the 
LBGs seem to match with the LAE contour well, indicating that they 
are residing in the same structure at $z=3.1$. 
The sight-lines are distributed over the LSPC in the Sb1 field, 
then, they are probing the ``intra-LSPC'' medium. 

Let us examine the H~{\sc i} absorption enhancement as a function of the
galaxy density. Although the number of sight-lines is not very large, we have
divided the sight-lines into three subsamples depending on the LAE overdensity
$\delta_{\rm LAE}$ reported by \cite{yam12} at the positions of the
sight-lines. Then, we have made their observer's frame
composites and measured the {\it excess} EWs. In this analysis, we have
kept the same wavelength range to measure the EW 
as that in Table~4. The results are summarized in Table~5. We find a weaker
{\it excess} EW for the lowest $\delta_{\rm LAE}$ subsample. However, 
it is not conclusive because the S/N remains low due to
the small number of sight-lines in the subsamples at the moment. 

\subsubsection{$z=2.98$ transparency peak}

This transmission peak is detected with a significance level of
$\ga3$\,$\sigma$. Remarkably, we have no LBG within the redshift range
of the transmission peak, while the expected number in a random
distribution is 8.2, which is calculated from the dashed
line in the top panel of Figure~\ref{DAcomp}. 
A Monte-Carlo simulation tells us that the
probability to have zero LBGs within the redshift range of $2.96<z<3.00$
is $\sim0.03$\%. 
This strongly suggests that this is a significant galaxy void. 
It is remarkable that this transparency peak at
$z=2.98$ indicating weaker Ly$\alpha$ absorption corresponds to the galaxy void 
besides the PC at $z=3.1$. 
It is also important that the Ly$\alpha$ absorption in
this void is not zero 
because the IGM optical depth would be 0.24 
for a mean transmission of 0.38 at the redshift (see eqs.~4, 8 and 14), 
indicating that there is substantial neutral hydrogen (or the LAF) 
even in a galaxy void, which is consistent with a result obtained 
in the low-$z$ Universe ($z\la0.1$) \citep{tej12}. 
Figure~\ref{z2p98peak} shows the distribution of the background sight-lines.

\subsubsection{Possible $z=3.24$ transparency peak}

As found in Table~3, this transmission peak is detected significantly in
the two $(S/N)_{\rm max}$ cases, but it is less than 3\,$\sigma$ in both
the $(S/N)_{\rm min}$ cases. Then, we consider this peak as a possible
detection. The number of LBGs within the feature is 2 against an
expectation of 6.3. A Monte-Carlo simulation predicts a probability
less than 5\% for 2 or a smaller number of LBGs in this redshift
range, and thus, it is a possible LBG void ($\sim1.6$\,$\sigma$). However,
we have to reserve a definite conclusion about the reality of 
this peak until more data become available.
Figure~\ref{z3p24peak} shows the distribution of the LBGs and sight-lines. 

\subsubsection{Possible $z=3.28$ absorption dip}

This absorption dip is significantly detected in the median composite
but not in the 3-$\sigma$ clipping average composite as found in
Table~3. Then, we consider this feature as a possible detection. On
the other hand, there is an overdensity of LBGs at
$3.26<z<3.31$ corresponding to the dip feature; the number of LBGs is 12 
against a random expectation of 6.2. A Monte-Carlo simulation tells us
the probability to have more than or equal to 12 LBGs within this
redshift range to be 2\%, corresponding to a significance level of
$\sim2$\,$\sigma$. Figure~\ref{z3p28dip} shows the
spatial distribution of LBGs which seem to cluster  
at the south-east quarter of the field. Unfortunately, the
IGM probing sight-lines do not distribute inside of the LBG structure but do
around it. This spatial displacement probably reduces the significance
of the IGM transmission feature, if real.
Clearly a much larger number of spectra is required to reveal 
the nature of this dip.

\subsection{Cross-correlation between galaxies and H~{\sc i} transmission}

We consider here the cross-correlation between the H~{\sc i}
transmission spectrum and the LBG redshift distribution
defined as follows to evaluate the degree of their  
synchronization. 
First, a spectrum expressing significance of the H~{\sc i} 
fluctuation can be defined as  
\begin{equation}
 \epsilon_\lambda \equiv \frac{\tilde{f_\lambda}-1}{\sigma_{\rm MC,\lambda}}\,, 
\end{equation}
where $\tilde{f_\lambda}$ is the normalized composite spectrum and
$\sigma_{\rm MC,\lambda}$ is the standard deviation in each wavelength
element estimated by our Monte-Carlo simulation. Next, we define a
spectrum describing the LBG overdensity significance as 
\begin{equation}
 \delta_{z(\lambda)} \equiv \frac{n_{z(\lambda)}^{\rm
  obs}-n_{z(\lambda)}^{\rm exp}}{\sigma_{{\rm LBG},z(\lambda)}}\,,
\end{equation}
where $n_{z(\lambda)}^{\rm obs}$ and $n_{z(\lambda)}^{\rm exp}$ are,
respectively, the observed and expected numbers of the LBGs in the
redshift $z$ interval corresponding to the wavelength element
$\Delta\lambda$ of the VIMOS setting used in our VI08 survey, and the
uncertainty $\sigma_{{\rm LBG},z(\lambda)}$ can be expressed as 
\begin{equation}
 {\sigma_{{\rm LBG},z(\lambda)}}^2 \approx {\sigma_{{\rm obs},z(\lambda)}}^2
  + {\sigma_{{\rm exp},z(\lambda)}}^2\,,
\end{equation}
where $\sigma_{{\rm obs},z(\lambda)}$ is given by a small number Poisson
statistics with the parameter $n_{z(\lambda)}^{\rm obs}$ \citep{geh86}
\footnote{The Poisson distribution with the parameter being a small number is asymmetric. 
\cite{geh86} gives upper and lower 84-percentiles. 
Since we need a single value for $\sigma_{{\rm obs},z(\lambda)}$, 
we use simple average values of the upper and lower percentiles. 
This choice determines the absolute value of the overdensity
significance and the resultant cross-correlation coefficient. 
To evaluate the significance of the observed 
cross-correlation coefficient, however, we do not need the absolute
value of the coefficients but need a relative comparison between the
observational and random ones. Therefore, this choice does not affect
our evaluation of the significance of the cross-correlation.} 
and 
$\sigma_{{\rm exp},z(\lambda)}\approx\sqrt{n_{z(\lambda)}^{\rm exp}}$ 
as also expected by the Poisson statistics, which we have confirmed by a
Monte-Carlo simulation. Finally, we define the cross-correlation
coefficient as
\begin{equation}
 \xi_{\rm cc} \equiv \frac{1}{n}\sum_{i=1}^{n} 
  \epsilon_{\lambda_i} \delta_{z(\lambda_i)}\,,
\end{equation}
where $\lambda_i$ is the $i$th wavelength element and $n$ is the total
number of wavelength elements used in the calculation. We restrict
ourselves to the wavelength elements where the number of the spectra
used in the composite is equal to or larger than 7 as in Figure~\ref{DAcomp}. 
The resultant coefficients are $-0.218$ and $-0.235$ for
the median and average composites, respectively. 

\begin{figure}
 \begin{center}
  \includegraphics[width=85mm]{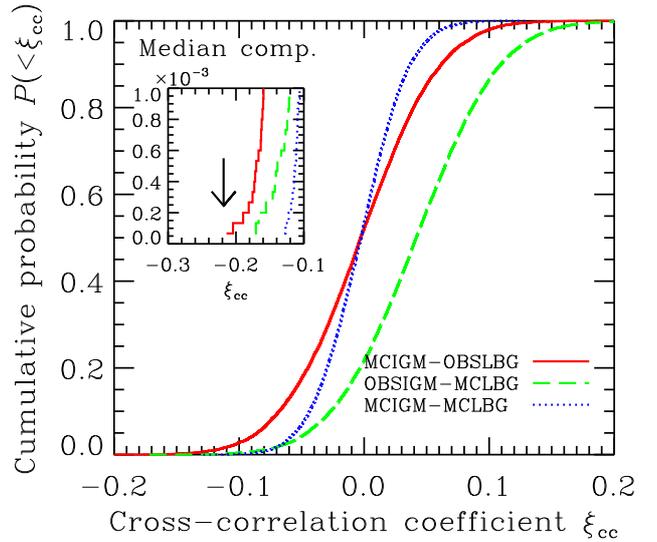}
 \end{center}
 \caption{Cumulative probability distribution to have a value of the
 cross-correlation coefficient defined by equation~(23) smaller than that
 in the horizontal axis for the median composite case. The solid line is
 the case with the combination of the Monte-Carlo IGM transmission and the
 observed LBG redshift distribution, the dashed line is the case of the
 combination of the observed IGM transmission and the Monte-Carlo LBG
 redshift distribution, and the dotted line is the case of the
 combination of both Monte-Carlo simulations. The inset is a zoom-in
 around the coefficient from the observed IGM transmission and the
 observed LBG redshift distribution indicated by the downward arrow.}
 \label{crosscor}
\end{figure}

For a comparison, we have performed an extensive Monte-Carlo simulation
of the IGM transmission and of the LBG redshift distribution. We have
15,000 mock IGM composite spectra generated by the Monte-Carlo
simulation as described in section 3.3. The simulation of the LBG
redshift distribution is based on the expected redshift distribution of
the randomly distributed LBGs as shown in the top panel of Figure~\ref{DAcomp}. 
Since we have 171 LBG redshifts, we randomly draw 171 redshifts from 
the expected function and repeat it 15,000 times. Then, we calculate the
cross-correlation coefficient, $\xi_{\rm cc}$, from these 15,000 sets of
the IGM transmission and the LBG redshift distribution. We try three
combinations of them: (1) the Monte-Carlo IGM and the real LBG redshift,
(2) the real IGM composite and the Monte-Carlo LBG redshift, and (3)
both data from the Monte-Carlo simulations. Figure~\ref{crosscor} shows the
cumulative probability function to have a coefficient $\xi_{\rm cc}$ 
smaller than the value in the horizontal axis for the 
median composite.  
We find that the observational coefficients noted at the
end of the previous paragraph are very rare in our
Monte-Carlo simulation: 
$<2\times10^{-5}$ and $7\times10^{-5}$ for the
median and average composites, respectively. Assuming a Gaussian
distribution, these values correspond to 
$>4.1$- and 3.8-$\sigma$ excesses, respectively.


Negative values of the cross-correlation coefficient mean an
anti-correlation of the IGM transmission and the LBG redshift
distribution, namely, an enhanced (reduced) Ly$\alpha$ absorption in a 
galaxy overdensity (underdensity). Since the negative values of the
coefficient obtained from the observed data are extremely difficult to
be explained with random distributions of the IGM and LBGs, 
we conclude that the IGM transmission significantly anti-correlates with
the LBG distribution.

From Figure~\ref{crosscor}, one can appreciate 
a bias towards positive values found in the combination of the observed
IGM transmission and the Monte-Carlo LBG distribution (the dashed line
in Figure~\ref{crosscor}). This is because the observed IGM fluctuation
$\epsilon_\lambda$ and the Monte-Carlo LBG overdensity
$\delta_{z(\lambda)}$ are both biased towards negative values. 
Since we calculate $\delta_{z(\lambda)}$ in the wavelength
pixel scale of our VIMOS spectroscopy and the numbers 
of LBGs in many pixels are then zero, resulting in negative $\delta_{z(\lambda)}$. 
In fact, the mean of $\delta_{z(\lambda)}$ from the observed LBG
distribution is also negative and very similar to those from the
Monte-Carlo simulation. On the other hand, the Monte-Carlo IGM
fluctuation gives a very symmetric distribution around zero. This is the
reason why we have obtained a median value of $\xi_{\rm cc}$ close to
zero with the Monte-Carlo IGM in Figure~\ref{crosscor}. The observed IGM
$\epsilon_\lambda$ tends to be negative: more Ly$\alpha$ absorption as
seen in the previous subsections.

We summarize correlation significance for each redshift range in Table~6. 
The high significance of the full redshift range described above is recognized 
to be a result of strong correlations mainly at $z=3.1$ and 2.98.  

\begin{table}
 \centering
 \begin{minipage}{140mm}
  \caption{A summary of significance of each correlation.}
  \begin{tabular}{lcc}
   \hline
   Redshifts & median $\sigma$ & average $\sigma$ \\
   \hline
   full $z$ range($z=2.46-3.55$)  & $>4.1$ &   3.8  \\
   $z2.98$ peak($z=2.97-3.00$)    &   3.7  & $>4.1$ \\
   $z3.10$ dip($z=3.07-3.12$)     &   3.9  &   4.1  \\
   $z3.24$ peak($z=3.22-3.26$)    &   2.3  &   2.3  \\
   $z3.28$ dip($z=3.26-3.31$)     &   2.3  &   2.0  \\
   all $z$ except $z3.10$ dip     &   3.1  &   2.8  \\
   all $z$ except $z3.10$ dip \& $z2.98$ peak & 1.9  & 1.5  \\
   \hline
  \end{tabular}\\
 \end{minipage}
 \label{CBstructureSN}
\end{table}

\section{Discussion I : \ \ The large-scale proto-cluster and surroundings }

\subsection{Cosmological characteristics of the large-scale proto-cluster}

In the beginning of the discussion, we estimate the total mass and a finding 
probability of the LSPC in the SSA22 field at $z=3.1$, 
which has induced the present spectroscopic survey, based on the LAE overdensity 
and an assumed bias parameter, amplitude of galaxy overdensities versus those
of matter. Here, we precisely define the LSPC as 
the area in the SSA22 Sb1 field where the local LAE number density
exceeds 1.5 times the mean value of the control fields, i.e., 0.204 LAEs
arcmin$^{-2}$ obtained in \cite{yam12}. The contours expressing the LSPC
area, the high density region (HDR) of the LAE is displayed in that
article. The HDR contains 259 confident LAEs defined in \cite{yam12} and
35 LABs including two gigantic ones by \cite{ste00}, which would be
considered to be progenitors of massive galaxies in the present
Universe, as well as around 50 LAAs, a number of LBGs and K-band
selected galaxies \citep{uch12}. So, the HDR is becoming to be called
the ``large-scale proto-cluster (LSPC)''. The FoVs of our VIMOS survey
have been set up to probe the LSPC.

The LAE number density of the LSPC is 0.58\,arcmin$^{-2}$, i.e., the overdensity 
$\delta$ of the LAE is $1.89\pm0.18$, which can be translated to the underlying 
matter overdensity of $\delta_M=0.99\pm0.25$, if we adopt the linear
bias parameter of $b_{LAE}=1.9^{+0.4}_{-0.5}$ for the LAE. 
This value was taken from \cite{gua10} (see also
\citealt{gaw07}) for $z\simeq3$ LAEs and would be reasonable 
compared to $b=2.6$ \citep{bie13} assumed in \cite{cuc14}
for the LBGs at $z=3$. The comoving volume of the LSPC 
is $0.92\times10^5$\,Mpc$^3$ indicating a radius of $R=27.8$\,cMpc
(comoving Mpc) of a corresponding spherical volume, for which a
1-$\sigma$ mass fluctuation is estimated to be $\sigma_M=0.12$ at
$z=3.1$ from the linear growth theory of CDM mass fluctuations with the
normalization $\sigma_8=0.81$, in the same manner as \cite{yam12}. This
means the probability of the LSPC amounts to
$8.3\pm2.1$\,$\sigma$, i.e., around $10^{-10}$. 
Although the volume of the LSPC is very large, the effect of the
gravitational evolution such as the gravitational
contraction of the massive structure should be taken into account to
obtain the probability accurately.
We evaluated it by the method of \cite{maw12} who used the log-normal probability 
distribution function (PDF) of underlying mass fluctuations as a reference. 
In the calculation of the PDF, the gravitational contraction of the structure 
and the effect of redshift distortion, the so-called Kaiser effect  
\citep{kai87} were taken into account. 
We estimated the PDF for mass fluctuations within the spherical region 
with a radius of $R=27.8$\,cMpc at $z=3.1$ (see Figure~11). From the function, we obtain the
appearance probability of $0.0023^{+0.0483}_{-0.0022} \%$ for the LSPC
with the mass overdensity of $\delta_M=0.99\pm0.25$.  
The probability is considerably larger than the estimate from the simple linear growth theory 
described above. It appears that the LSPC with the extremely large total mass of  
$\approx0.90\times10^{16}$\,M$_{\odot}$ has already begun the
gravitational contraction at $z=3.1$.


\begin{figure}
 \begin{center}
  \includegraphics[width=80mm]{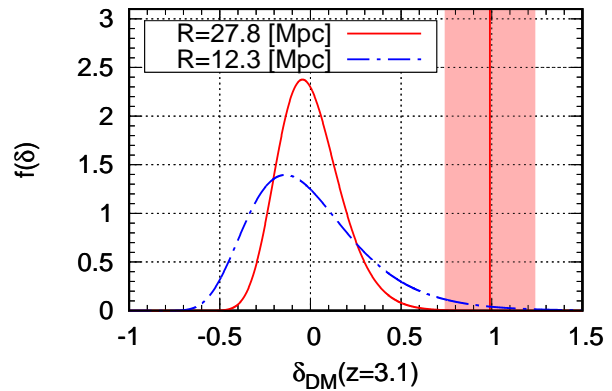}
 \end{center}
 \caption{The probability distribution function (PDF) of the dark-matter mass
 overdensity $\delta_{\rm DM}$ at $z=3.1$, as a result of the
 cosmological evolution of the initial Gaussian mass fluctuation, within
 a sphere having the radius of 27.8 comoving Mpc is shown by the solid
 curve. The dot-dashed curve is the case with a radius of 12.3 comoving
 Mpc for a comparison. The vertical solid line and shaded region are,
 respectively, $\delta_{\rm DM}$ and its uncertainty for the $z=3.1$
 LSPC in the SSA22 field.}
 \label{PDFdist}
\end{figure}

\subsection{H {\sc i} transparency peaks close by the large-scale proto-cluster}

We have detected two significant correlations between LBGs and H~{\sc i}
transmission with 4\,$\sigma$ or more at $z=3.10$ and 2.98, together
with two possible ones at $z=3.24$ and 3.28 with about 2\,$\sigma$
significance, in the previous section. We call redshift coincidences
between LBG-HDR and the absorption dip in Figure\,\ref{DAcomp},
``Counter-Balance structure 1, CB1'' as well as the ones between the 
LBG low density region (LDR) or void and transparency peak in the figure, ``CB2'',
respectively. In this subsection, we discuss CB1 at $z=3.10$ 
and CB2 at $z=2.98$ having sufficient significance, qualitatively, and
briefly mention the simultaneous appearance of two transparency peaks at $z=2.98$ and 3.24.

The CB1 is not too difficult to be understood, because high density LBGs
and their H~{\sc i} halos absorb photons at the Ly$\alpha$ wavelength
with high probabilities. H~{\sc i} gas proper to the LBG cluster may
also contribute to make a dip, as suggested in \cite{maw17}.  
On the other hand, the CB2, the high transparency peak
seen in the transmission spectrum at the LBG void or LDR, is not so easy
to be interpreted. As the simplest interpretation, in a void/LDR,
absence or underdensity of LBGs with H~{\sc i} halos would cause such
transparency in H~{\sc i}. It is valid, if the
Ly$\alpha$ depression is caused mainly by LBGs and their H~{\sc i}
halos. However, this picture does not seem to be true. 
For example, we can find LBG-LDRs without transmission peaks at $z=2.80$
and 2.90 in Figure~\ref{DAcomp}.
For the former LDR, the number of LBGs detected at redshifts between 
$z=2.79$ and 2.84 is only two, 
against 9.7 LBGs expected in the uniform distribution 
shown by the dashed line in Figure~\ref{DAcomp}~(a). 
A Monte-Carlo simulation gives the probability of around 
0.3\% to have two or a smaller number of LBGs in this
redshift range, implying the $z=2.80$ is an LBG void with $\sim3\sigma$ significance. 
However, any transparency peak is not seen at all at
the redshift in the transmission spectrum in Figure~\ref{DAcomp}~(b), despite a lot
of sight-lines. The narrow LDR at $z=2.90$, where one
LBG is found at  $2.89\leq z\leq 2.91$ whose probability is estimated at
5.6\% ($\sim1.6\sigma$), also exhibits no transparency
peak. This could mean that a significant fraction of HI absorption in
these regions is not directly associated to galaxy halos
(e.g. \citealt{tej12} for a similar conclusion at low redshifts).

At redshifts lower than $z=2.8$, from $z=2.75$ down to 2.55, it is
difficult to discuss the LBG redshift distribution and identify
voids/LDRs with sufficient significance because the LBG detection
efficiency in their redshifts is low as indicated by the dashed line in
Figure~\ref{DAcomp}~(a). The CFHT $u^*$-band used in our LBG selection
to detect dropout phenomena has a relatively long central wavelength and
the lower bound of the selection redshift becomes relatively high. On
the other hand, the VIMOS sensitivities are not too low to measure the
spectra at wavelengths between 4300\,\AA\ and 4600\,\AA, corresponding
to $z=2.55$--$2.75$ in Ly$\alpha$, and the composite transmission
spectrum consists of a lot of sight-lines as shown in
Figure~\ref{DAcomp}~(c). So, in this redshift range, we can perform
H~{\sc i} transmission measurements with sufficient  
significance. According to the cosmological simulations such as the
Millennium simulation \citep{spr05}, there should be
several LBG voids/LDRs with the sizes of dozens of cMpc at redshifts
between $z=2.55$ and 2.75 (see also \citealt{sta15b}).
We call them, which are expected at those redshifts but difficult to be
recognized in our LBG survey, potential voids/LDRs. It is important that
the observed voids/LDRs at $z=2.80$ with $\sim3\sigma$
and $z=2.90$ with $\sim1.6\sigma$ as well as potential ones at
redshifts less than 2.75 make no transparency peaks at
all as seen in Figure~\ref{DAcomp}~(b).  
Therefore, absence or underdensity of LBGs in the ordinary void itself 
does not seem to cause such a prominent transparency peak found at $z=2.98$. 

To understand the generation of the transparency peak
at $z=2.98$, we put an attention to the structure $\sim100$~cMpc away along the line-of-sight: 
LSPC at $z=3.10$ having large overdensities of LBGs and LAEs studied in the previous subsections.  
The LSPC defined in the SSA22 Sb1 field is considered to have the total mass of 
$\approx0.90\times10^{16}$\,M$_\odot$ including the overdensity mass of
$0.45\times10^{16}$ M$_\odot$ ($\delta_M=0.99$) 
under the assumption of a bias parameter of 1.9 for LAEs as discussed above. 
This extremely large overdensity will attract the matter from the
regions surrounding the LSPC, decrease the matter density there, and
accelerate the expansion of their space by so-called ``tidal force''. If
some regions of the surroundings have already been low density compared
with the mean at an early epoch as a result of the hypothesized quantum
fluctuation during the inflation, the region will effectively grow into
a {\it real void}, i.e., devoid of galaxies {\it and} H~{\sc i} gas,
under the strong gravity of the ``nearby'' massive LSPC, after the
$t_{eq}$, the moment when the matter energy density just exceeds the
radiation one. Even if the region did not exhibit a very low density
fluctuation with a high $\sigma$, the tidal force induced by the
``nearby'' massive structure would accelerate a growth of the region
into a sufficient void with less LBGs. Such an extended space will also
cause a low LAF density on sight-line. In this way, the CB2 structure
seen at $z=2.98$, the LBG void associated with the H~{\sc i}
transparency peak, would only be formed with a help of
a nearby massive structure such as the LSPC at $z=3.1$. We can test this
hypothesis by finding more extreme high-density regions and
hypothesizing that there should be voids of type CB2 nearby these
structures too.

Moreover, in the panoramic survey for seven S-Cam FoVs, Sb1-7 in SSA22,
very interestingly, more HDRs, defined as the area where the local LAE
number density exceeds 1.5 times of the mean of the control fields,  
are found in Sb2-7 besides the LSPC in Sb1. The contours of their HDRs
are displayed in the panoramic sky map in \cite{yam12}. The total mass
of these HDRs in the panoramic survey amounts to 
$3.2\times10^{16}$\,M$_\odot$ including the LSPC in Sb1, of which the
overdensity mass is estimated to be $1.5\times10^{16}$\,M$_\odot$ with
$b=1.9$ for LAEs at $z=3.1$.  
This huge overdensity mass will work as a source of the ``tidal force'' 
as discussed in the following.

Here, we have to consider that these overdensity masses obtained from
the NB survey are limited to the space sliced by the NB filter whose
thickness is 58\,cMpc. So, it is only a part of the entire overdensity
mass responsible to the tidal force for the space around $z=2.98$,
although it is already huge: $1.5\times10^{16}$\,M$_\odot$. We need the
three-dimensional structures and overdensity map at least for the region
within a radius of 100\,cMpc around the LSPC at $z=3.1$. Future
wide-field spectroscopic surveys will provide the entire overdensity
mass in this region to obtain the exact tidal force and accurately
calculate the expansion of the surroundings of the LSPC.

In addition, a smaller concentration of LBGs between
$z=2.91$ and 2.96 can also contribute to the local space expansion at
around $z=3.00$. The structure, a modest HDR with a mean redshift of
$z=2.93$, has the overdensity of $\delta_{LBG}=0.5$ compared with the
dashed line for the uniform distribution in Figure~\ref{DAcomp}~(a),
which can be converted into the mass overdensity of $\delta_M=0.2$ by
applying $b_{LBG}=2.6$ \citep{bie13} previously used. 
Note that \cite{ste98} found a damped Ly$\alpha$ (DLA)
system at $z=2.93$, suggesting the reliability of the overdensity of
this modest ``HDR''. If we assume that the modest ``HDR'' has a
spherical form with a diameter of 50\,cMpc corresponding to the redshift
interval of $dz=0.050$ described above, the comoving volume becomes
$6.5\times10^4$\,cMpc$^3$. Then, the overdensity mass of the modest HDR
with $\delta_M=0.2$ turns out to be $0.06\times10^{16}$ M$_\odot$, which
is about one-20th of the LSPC overdensity mass at $z=3.10$. So, the
gravity by the two sources is almost balanced at $z=2.96$, the lower
redshift edge of the $z=2.98$ void. On the other hand, the gravitational
force by the LSPC dominates the other edge at $z=3.00$. In this way, the
LSPC and probably associated HDRs around it should effectively expand
the space between $z=2.96$ and 3.00, to make both the LBG void and
transparency peak at the redshift, i.e., CB2 structure.  

Likewise, the transparency peak at $z=3.24$ would be
induced by the LSPC at $z=3.10$ together with a modest HDR at $z=3.28$
seen in Figure~\ref{DAcomp}~(a), just behind the peak, with the same  
mechanism of the tidal expansion as the $z=2.98$ peak formation.

\subsection{Extended H~{\sc i} halo of LBGs}

Using a large sample of foreground-background galaxy
pairs, \cite{ste10} revealed that LBGs have large H~{\sc i} halos from
the composite spectra at the rest-frame of foreground galaxies. 
The H~{\sc i} halo extends to 0.3\,proper\,Mpc (pMpc) corresponding to 
1.2~cMpc at redshift 3 and 30\,arcsec in the angular scale.  
For example, the rest-frame equivalent width of the H~{\sc i} amounts to around 
0.3\,\AA\ at the impact parameter $b=0.3$ pMpc. It can be said that their findings 
have changed the picture of galaxies at high redshifts.  
For QSO environments, \cite{pro13,pro14} presented remarkable radial
profiles of H~{\sc i} and metal absorption of the 
circum-galactic medium of $z\sim2$
massive galaxies hosting QSOs by using their sample of QSO pairs.
Their findings also have changed the picture of QSO environment. 

Following the studies on LBG halos by Steidel et al.,  
\cite{rak12} and \cite{rud12} have found H~{\sc i} ``halos'' 
which extend to surprisingly large distances of around 2\,pMpc, 
by using pairs of galaxy and background-QSO. Such an extension of 
a single galactic ``halo'' is quite strange, because 2\,pMpc corresponding 
to 8~cMpc at redshift 3 is a typical scale of cluster of 
galaxies, i.e., it means that each LBG has an H~{\sc i} ``halo'' of the same 
extension as clusters of galaxies. 

To understand the large H~{\sc i} ``halo'', it is important 
to look into the two-dimensional H~{\sc i} absorption
map in the transverse and sight-line distances 
presented by \cite{rak12} (see also \citealt{tur14}). 
H {\sc i} at the impact parameter $b<0.13$\,pMpc shows the 
Finger-of-God, suggesting its virial motion in a galaxy, and H~{\sc i}  
at $b>0.13$\,pMpc exhibits the Kaiser effect \citep{kai87}. 
The map indicates that H~{\sc i} at $b<0.13$\,pMpc clearly belongs to
the host LBG, and H~{\sc i} at $b>0.13$\,pMpc can be interpreted as the
falling cool gas to the LBG as pointed out in \cite{rak12,rak13}.
If the LBG belongs to a cluster of galaxies, the falling cool H~{\sc i} would 
be supplied by the cluster. If it is a field galaxy, the H~{\sc i} is probably
supplied by the intervening cosmic web around the LBG.
Here, it is difficult to understand at present whether the falling 
H {\sc i} gas is galactic medium or intergalactic. However, the virial
H~{\sc i} gas of  $b<$0.13\,pMpc clearly belongs to the LBG.

The PC at $z=3.1$ in the SSA22 field has a large LBG
overdensity, $\delta=3$ for our VIMOS survey area. Therefore, the virial
H~{\sc i} with $b<$0.13\,pMpc around the LBGs, which distribute in the
cluster with a high density, may reproduce the observed excess
absorption of $EW_0=-1.7$\,\AA.  
In this case, there is no neutral hydrogen proper to the PC, 
which suggests that LBGs in the cluster are in a stage of the lack of fuel. 
On the other hand, there should be an H~{\sc i} supply to LBGs, if the
virial H~{\sc i} can produce only a part of $EW_0$ of $-1.7$\,\AA. 
In a future work, we will use a Monte-Carlo method to study whether 
the virialized H~{\sc i} around LBGs in the PC is able to produce 
the observed $EW_0$ or not, taking into account contributions of faint LBGs.

\section{Discussion II : \ \ Inhomogeneous structures accompanied with AGNs } 

It is generally recognized that LBG surveys using the $U$-dropout method
are also effective for QSO/AGN detections at $z\sim3$, because both
spectra are usually similar in respect of the Lyman break in the
$U$-band \citep{bie11}. In fact, we detected 4 AGNs in VI08 and
\cite{ste03} also found two AGNs in this field in their $z\sim3$ LBG
survey. In addition, we identified 5 AGNs in our VIMOS survey carried
out in 2006, VI06. It was 0.5 magnitude shallower than VI08 because of
about half of the integration time of VI08
\citep{kou11}, and was not deep enough to detect LBGs having UV
magnitudes fainter than around 25 AB. Therefore, we have
not used VI06 data in the previous sections dealing with LBG spectra
down to 25.4 AB. VI06 was a pilot observation for the VI08 LBG survey. 

On the other hand, VI06 is deep enough to observe AGNs with UV continuum
magnitudes brighter than 24.5 AB. LBG selection criteria of VI06 were
similar to those of VI08 and both expected redshift histograms are thus
similar: about uniform but slowly changing efficiency between $z=2.7$
and 3.5 with a peak at $z=3$ (see Figure~\ref{zdist}~[a]).
 
\begin{figure}
 \begin{center}
  \includegraphics[width=70mm]{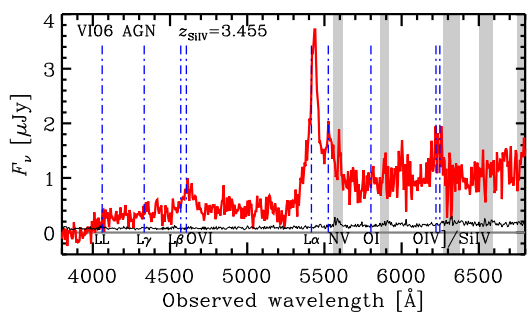}
 \end{center}
 \caption{One-dimensional spectrum of the AGN at $z=3.455$. 
 The Lyman limit system is seen at $z=3.44$. 
 The grey shaded regions indicate noisy wavelengths due to night
 emission lines. The vertical dot-dashed lines indicate
 some emission/absorption features.}      
 \label{AGNz3.455spec}
\end{figure}

In the following subsections, we discuss the nature of the AGN
distribution and its number density, especially, the relation between
AGN and LBG distributions at $z=3$--4 in the SSA22 field, using the AGNs
from the three surveys described above. In Table~7, we present
coordinates, redshifts, $R$ magnitudes, and survey names of the 11 AGNs.  
We show the spectrum of an AGN from the VI06 survey with $z = 3.455$ as
an example in Figure~\ref{AGNz3.455spec}.

 
\begin{table*}
 \centering
 \begin{minipage}{150mm}
  \caption{List of 11 AGNs found in Steidel et al. (2003) and our VIMOS surveys.}
  \begin{tabular}{lccccc}
   \hline
   Object & RA (J2000) & DEC (J2000) & Redshift & $R$ [AB] & Survey \\
   \hline
   z2.42 & 22:17:04.87 & $+$00:09:40.38 & 2.42 & 24.53 & VI08 \\
   z2.50 & 22:18:31.36 & $+$00:20:22.67 & 2.503 & 25.24 & VI06 \\
   z3.084 & 22:17:36.51 & $+$00:16:22.9 & 3.084 & 21.61 & Steidel et al. \\
   z3.091 & 22:17:16.23 & $+$00:17:44.88 & 3.100 & 24.32 & VI06 \\
   z3.104 & 22:17:09.62 & $+$00:18:01.04 & 3.112 & 24.41 & VI06 \\
   z3.112 & 22:17:12.60 & $+$00:29:02.76 & 3.110 & 23.82 & VI08 \\
   z3.132 & 22:17:06.75 & $+$00:26:41.27 & 3.132 & 23.99 & VI08 \\
   z3.353 & 22:17:22.26 & $+$00:16:40.41 & 3.353 & 21.23 & Steidel et al., VI06 \\
   z3.425 & 22:18:04.14 & $+$00:19:46.88 & 3.425 & 24.62 & VI08 \\
   z3.455 & 22:17:51.34 & $+$00:20:36.66 & 3.455 & 22.87 & VI06 \\
   z3.795 & 22:17:05.37 & $+$00:15:14.25 & 3.801 & 22.01 & VI06 \\
   \hline
  \end{tabular}
 \end{minipage}
\end{table*}

\subsection{Extreme overdensity of AGNs at $z=3.1$}

 
\begin{figure*}
 \begin{center}
  \includegraphics[width=120mm]{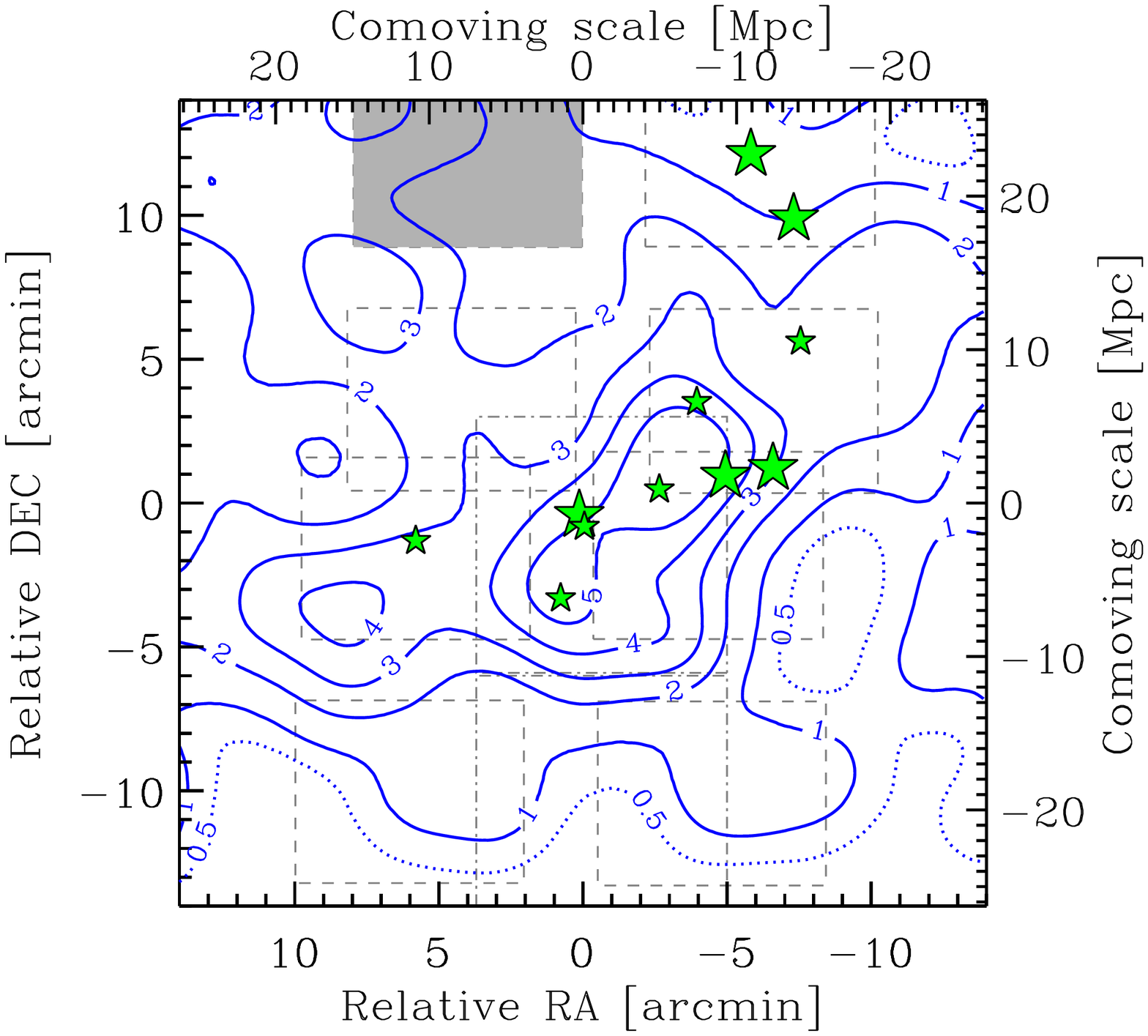}
 \end{center}
 \caption{Sky distribution of 11 AGNs at $z=3.084$--3.132. Large green
 stars are the AGNs with $R<24.5$, and small green stars are ones with
 $R>24.5$. The contours, gray lines and shaded area are
 the same as in Figure~\ref{z3p10dip}.}
 \label{AGNz3p1skymap}
\end{figure*}

It is remarkable that there are 5 AGNs with $R<24.5$ in a narrow
redshift range, $z=3.084$--3.132, where the LSPC and its envelope
lie. We show the sky map of their AGNs with large green stars in
Figure~\ref{AGNz3p1skymap}. Three of the five AGNs exist in the LAE
density peak area.

The sky area they occupy is around $15' \times 8'$, i.e.,
$27\times15$\,cMpc$^2$. The redshift difference of $dz=0.05$ corresponds
to about 45\,cMpc. So, the number density of the AGN cluster is 5
AGNs/$2\times10^4$\,cMpc$^3$. On the other hand, only 0.1 AGNs with
$R<24.5$ are expected in this volume from the QSO/AGN luminosity
function (LF) for $z=3.2$ by \cite{mas12}. Even if we take a comoving
volume of the entire LSPC, $1\times10^5$\,cMpc$^3$, the expected number
of AGNs is 0.5. 
 
It is noteworthy that about 4 AGNs with $R<24.5$ are expected from their
LF for our effective survey volume of $1\times10^6$\,cMpc$^3$, i.e.,
$z=2.7$--4.0 $\times$ 322 arcmin$^2$. This is consistent with our three
AGNs of $R<24.5$ detected at $z=2.7$--4.0 with the LBG selection
criteria, except for the LSPC redshift. The extreme AGN concentration
probably caused by the LSPC is a very interesting phenomenon to be
intensively studied, in conjunction with overdensities of LBGs, LAEs,
LABs and other kinds of objects in this field.
 
We also plot 6 $z=3.1$ AGNs fainter than $R=24.5$ from \cite{mic17} with
small stars in Figure~\ref{AGNz3p1skymap}. It is very interesting that
the 11 AGNs in total form a filamentary structure along the LAE density
peak. This structure at $z=3.1$ could indicate important characteristics
and dynamics on the formation mechanism and activities of AGNs. Future
studies of the relation between the LSPC and AGNs as well as LBGs, LAEs,
LABs, and so on, will offer us new insights on structure formation in
the early Universe.
 
In contrast to our SSA22 survey, for example 
\cite{bie11}, in which about 1000 LBGs are identified in 10 times larger
volume than ours, finds no remarkable AGN concentrations as well as high
density peaks of LBGs like the $z=3.1$ LSPC.
 
In the following sections, we discuss the other 4 AGNs found behind the
LSPC in the SSA22 field, apart from $z=3.1$. Especially, correlations
between AGNs and LBGs are considered.
  

\subsection{``Sheet-like'' structure of LBGs with an AGN at $z=3.353$}

\begin{figure*}
 \begin{center}
  \includegraphics[width=120mm]{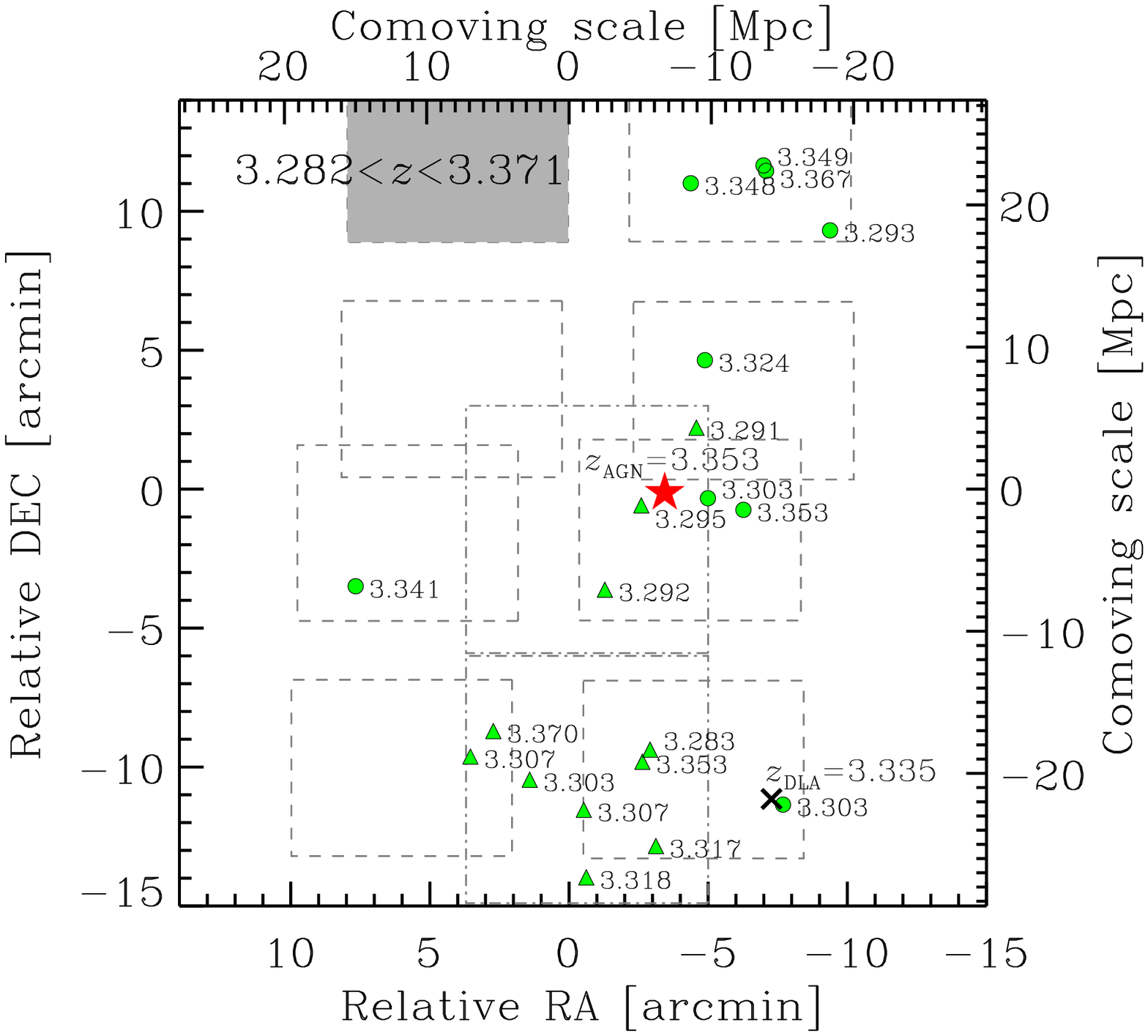}
 \end{center}
 \caption{Sky distribution of 20 LBGs at $z=3.28$--3.37. The green
 circles and triangles are the LBGs detected in the VI08 and Steidel et
 al.~(2003) surveys, respectively. 
 The numbers near the symbols indicate the redshift. 
 The AGN at $z=3.353$ is shown by the  
 star symbol and the DLA at $z=3.335$ is shown by the
 cross. The gray lines and shaded area are the same as in 
 Figure~\ref{z3p10dip}.}
 \label{LBGz3p35skymap}
\end{figure*}

The AGN at $z=3.353$ is detected and spectroscopically measured by both
\cite{ste03} and VI06. The AGN has a DLA at $z=2.93$ \citep{ste03}. In a
viewpoint of AGN-LBG correlation, we take notice of a modest LBG concentration 
consisting of 20 LBGs at $z=3.28$--3.37 around the
AGN's redshift in Figure~\ref{DAcomp}~(a). We show the sky map of
these LBGs in Figure~\ref{LBGz3p35skymap}. 
Interestingly, the LBGs, except for the east-most one, 
form a filamentary structure in the redshift slice of $z=3.28$--3.37, 
which extends along the north-south direction.  
The AGN lies at the center of the filament. Although the 
LBG redshift distribution for the entire FoV of
322\,arcmin$^2$ in Figure~\ref{DAcomp}~(a) may not show a strong
evidence for an ``LBG HDR'', the filamentary structure 
shown in Figure~\ref{LBGz3p35skymap} supports the reality of the LBG
high density.


Redshifts of the LBGs in the filament distribute almost
uniformly between $z=3.28$ and 3.36, therefore the filamentary HDR in the
two-dimensional sky map may be interpreted to be a sheet-like structure
in the three-dimensional space having a narrow gap at $z=3.33$--3.34
(see Figure~\ref{DAcomp}~[a]).
In addition, \cite{maw16} reported a DLA at $z=3.335$
probably associated with this sheet-like structure, further enhancing
the reality of the structure.
 
We discuss the overdensity and appearance probability of the HDR. The
``sheet-like'' HDR would be considered to have a mean width of 7\,arcmin
(13\,cMpc) and a length of 27\,arcmin (50\,cMpc) in the sky plane
represented in Figure~\ref{LBGz3p35skymap}, as well as the thickness of
75\,cMpc corresponding to the redshift interval of the 19 LBGs, $dz=3.370-3.283=0.087$. 
Thus, the comoving volume of the ``sheet-like'' HDR is $5 \times 10^4$ cMpc$^3$, 
in which 6.0 LBGs brighter than $R=25.4$ AB are expected when LBGs
distribute uniformly, as the dashed line shows in Figure~\ref{DAcomp}~(a). 
In this way, the HDR has a number overdensity 
$\delta_{LBG}=2.2\pm0.7$, which results in a mass overdensity 
$\delta_{M}=0.83\pm0.28$ when the bias parameter of $b=2.6$ \citep{bie13}  
for LBGs is employed. On the other hand, the $1\sigma$ fluctuation of 
the dark matter for this volume at $z=3.33$ is estimated to be 0.124,
according to the formula described in \cite{yam12}. Then, this LBG HDR
associated with the AGN at $z=3.353$ is a rare event with the appearance
probability of $7\pm2\sigma$. 


\subsection{Pair of AGNs around a dense H~{\sc i} cluster at $z=3.453$} 

We detected a pairwise AGN at $z=3.455$ and 3.425 
with an angular distance of about 3 arcmin, whose spectra are shown in
Figures~12 and 15 respectively. The former is a Type I AGN with
$R=22.87$, and the latter is a Type II AGN with $R=24.62$, whose
spectrum is very similar to the composite one for narrow-line AGNs in \cite{hai11}.
We express the redshift of the pair with $z=3.455$. The pair AGN
does not seem to have any clusters of LBGs with the  
$R$-band magnitudes brighter than 25.4 AB around their redshifts in
Figure\,\ref{DAcomp}~(a), although the detection efficiency for the LBGs
is not so high at the redshifts. So, the AGNs may be
interpreted as field objects independent of galaxy clustering.   
If our targets of the VI08 survey were only ``objects'' like galaxies in 
the ordinary survey, the pair would be misidentified as an isolated one
in space. However, our survey can examine neutral hydrogen gas besides
``objects'', as discussed in the previous sections. So, we investigate
H~{\sc i} absorption dips around the pair of AGNs imprinted in DA
ranges of spectra of background objects. 
Especially here, we search for H~{\sc i} gas clustering in the area within 
an angular radius of 10 arcmin from the center of the pair, which corresponds 
to 20 cMpc at $z=3.455$, a typical size of the PC at high redshifts.

 
\begin{figure}
 \begin{center}
  \includegraphics[width=70mm]{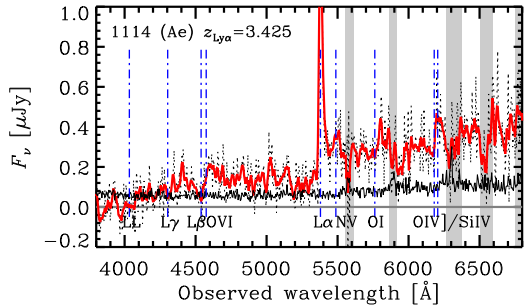}
 \end{center}
  \caption{One-dimensional spectrum of the AGN at $z=3.425$, which shows
 narrow emission lines. The gray shaded regions indicate 
 noisy wavelengths due to night emission lines. The vertical dot-dashed
 lines indicate some emission/absorption features.}      
 \label{AGNz3.425spec}
\end{figure}


\begin{figure*}
 \begin{center}
  \includegraphics[width=120mm]{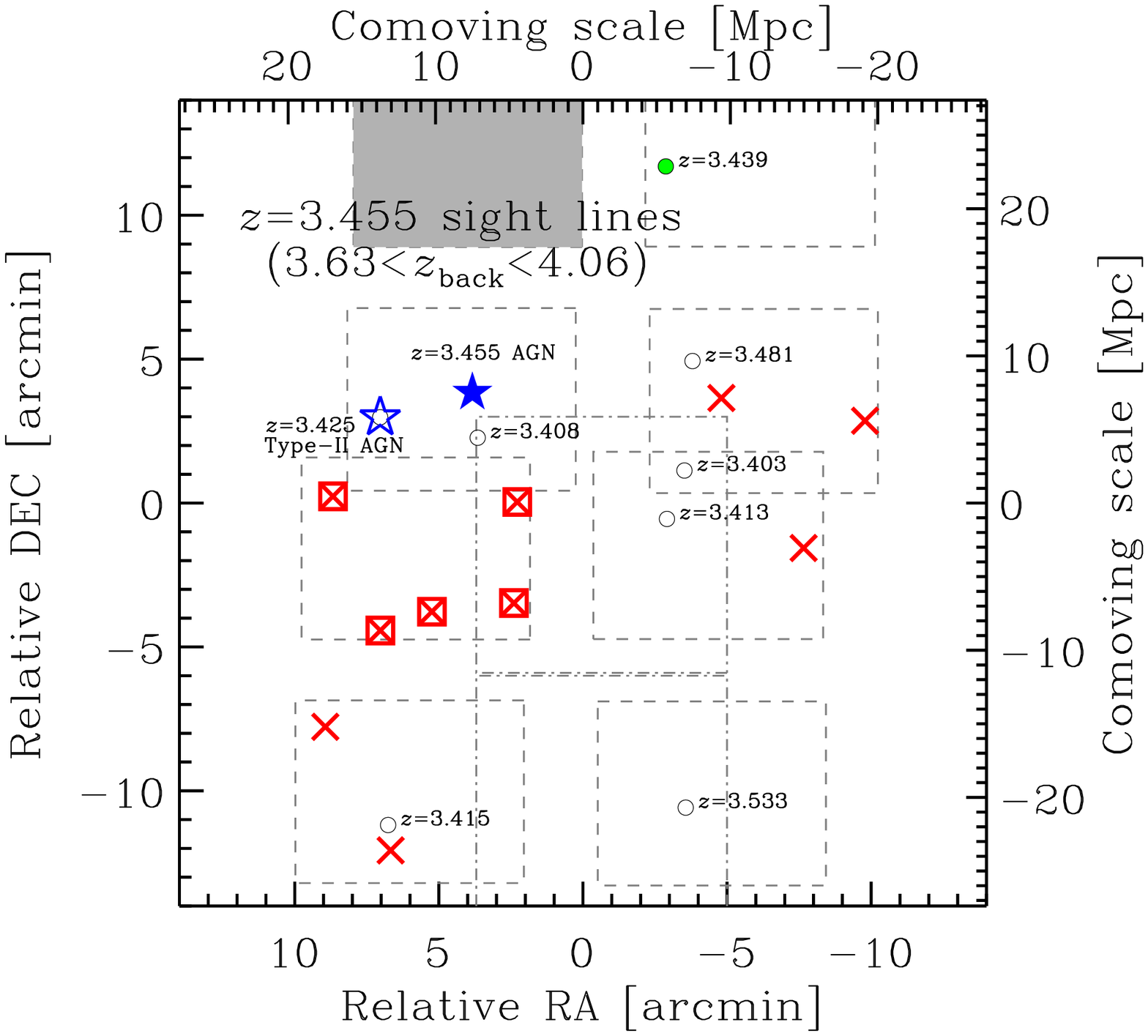} 
 \end{center}
 \caption{Sky distribution of 10 sight-lines (cross marks) in the plane
 at $z=3.455$. The 5 sight-lines showing H~{\sc i} absorption
 enhancement are indicated by the square enclosing crosses. The Type-I
 (Type-II) AGN at $z=3.455$ (3.425) is shown by the filled (open) blue
 star. The circles indicate the positions of LBGs at $3.40<z<3.60$ with
 their redshifts and one green filled circle is an LBG in the redshift
 range of the H~{\sc i} absorption enhancement. The gray lines and shaded area 
 are the same as in Figure~\ref{z3p10dip}.}
 \label{SLz3p455skymap}
\end{figure*}
 
\begin{figure}
 \begin{center}
  \includegraphics[width=70mm]{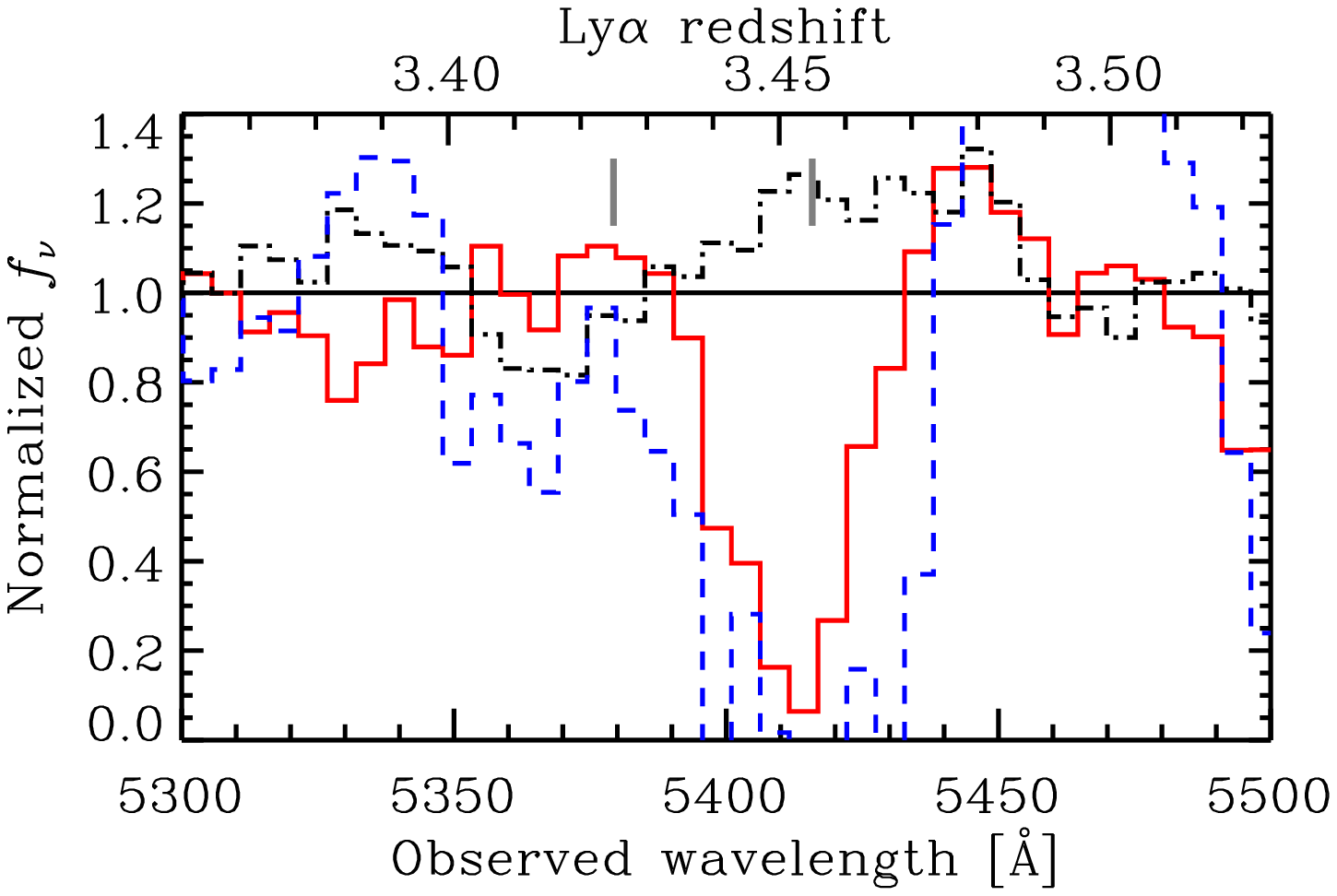}
 \end{center}
 \caption{Observer's frame composite spectrum of the LBGs behind of the
 pair AGN at $z=3.455$. The solid histogram shows the
 median composite of the 5 background LBGs south of the pair AGN 
 (those indicated by the square enclosing crosses in Figure~16). The composite 
 spectrum of the other 5 background LBGs in Figure~16 is shown by the dot-dashed 
 histogram. The dashed histogram shows the normalized DA spectrum of the nearest 
 sight-line to the Type-II AGN. The two vertical gray ticks show the redshifts 
 of the AGNs.} 
 \label{compospec_z3.455}
\end{figure}

In Figure~\ref{SLz3p455skymap}, we show the sky map of the pair AGN
together with sight-line positions of 10 background objects having
redshifts of $z=3.63$--4.03, whose DA ranges cover Ly$\alpha$ at
$z=3.425$--3.455. There are 5 sight-lines forming  
a ``cluster'' with a diameter of about 8 arcmin in the south part of the
pair (the squares enclosing crosses in 
Figure~\ref{SLz3p455skymap}). We have made a composite spectrum of the  
5 sight-lines and show the result in Figure~\ref{compospec_z3.455},
which clearly shows a deep absorption dip with the rest-frame EW of
about $-5$~\AA\ at 5415~\AA, corresponding to $z=3.453$, if Ly$\alpha$
absorption is assumed.
In contrast, the composite of the remaining 5 sight-lines shows no dips
at the redshift as seen in Figure~\ref{compospec_z3.455}. 
Two of the 5 sight-lines in the ``cluster of sight-lines'' have very deep absorption dips 
like sub-DLA/Lyman limit systems (LLSs) at $z=3.453$. Also the other 3 sight-lines 
have considerably significant dips at the redshift, indicating the dense H~{\sc i} gas 
at $z=3.453$ extends over the entire 5 sight-line area with a diameter of at least 8 
arcmin. We call the area a dense H {\sc i} cluster (DHC).



Interestingly, the spectrum of the $z=3.455$ AGN seems to have 
an LLS at 4050\,\AA, i.e., $z=3.441$ in Figure~\ref{AGNz3.455spec}, 
indicating that there is plenty of H {\sc i} gas also around the AGN. 
Therefore, the DHC found in the composite spectrum would not only cover the 5 sight-lines 
but also extend to cover the sight-line towards the $z=3.455$ AGN.  
Here, we notice a slight difference between the Type-II AGN redshift,
$z=3.425$ and the composite dip one, $z=3.453$ in Figure~17. However,
the nearest sight-line from the $z=3.425$ AGN exhibits a considerable
absorption feature down to $z=3.40$ including the AGN redshift as shown
by the blue dashed line in Figure~17, although the statistical
significance is not high enough. Thus, the DHC has an angular diameter
of at least about 10 arcmin and would have a three-dimensional structure
extending to lower redshift around a sky position of the $z=3.425$ AGN
to include both AGNs as its members. 

If the DHC spans the area of a 10 arcmin (or 20 cMpc) diameter
(roughly $1/3$ of the entire observing field) and the redshift range of
$z=3.420$--3.470, the expected number of the $R<25.4$ AB LBGs is 1.3
against no LBG in the area. Therefore, the pair AGN does not 
correlate with any LBG HDR. Nevertheless, it seems to be 
strongly associated with the DHC of a 20 cMpc scale. Such a relation is
very interesting, even strange, and may be a new type of AGN-matter 
correlation. 
The rich H~{\sc i} gas in the 20 cMpc scale cluster should contribute to
pair AGN generation and keeping their activities. Mechanism to feed the
fuel to the AGNs probably having the super-massive black-holes (SMBHs),
in the DHC of a 20 cMpc scale, should be intensively studied and revealed.

In the three-dimensional LBG distribution of  
\cite{bie11}, a small fraction of AGNs seems to exist in LBG LDRs or
voids. It usually means that their AGNs are isolated in space, i.e., 
field objects. However, they may exhibit some correlations with dense
H~{\sc i} gas, as discussed above for the SSA22
field. Surveys for them are important and interesting to understand
mechanisms of formation and activity of AGNs. 

Comparing the present result in this subsection with the work by
\cite{cuc14} is interesting. They found a very deep H~{\sc i} absorption
dip of a rest-frame EW of $-10.8$~\AA\ with the comoving volume of
$13\times15\times17$ cMpc$^3$ at $z=2.895$ in the COSMOS field through
the VIMOS Ultra-Deep Survey (VUDS). Its size is similar to that of our DHC, 
about $20\times20\times40$ cMpc$^3$. Also, its rest-frame EW is comparable 
with our $-5$~\AA. The great disparity is the counterpart. The absorption dip 
from \cite{cuc14} is associated with a large LBG overdensity of $\delta_{LBG}\sim12$. 
However, our deep H~{\sc i} absorption dip does not show such an overdensity peak of
LBGs as a counterpart in the present statistics. 

We stress that the number density of LBGs is very large compared to
AGNs, i.e., the comoving density of the LBGs with $R<25.4$ is about 
60 times larger than that of the AGNs with $R<25.4$ from their LFs at 
$z\sim3.5$ \citep{mas12}. In spite of such popular
objects, any LBGs with $R<25.4$ are not yet detected in the DHC which
was found triggered by the presence of the pair of 
AGNs. As discussed in this paper, the DHCs, the regions showing strong
H~{\sc i} absorption dips in the composite spectrum, have a tendency to
show significant correlations with LBG overdensities, as the LSPC at
$z=3.1$ for the H~{\sc i} absorption dip at the same redshift and the
overdensity of LBGs at $z=3.26$ - 3.31 discussed in subsection 4.2.4 for
the H~{\sc i} dip at $z=3.28$ in the SSA22 field, together with the LBG 
sharp peak found at $z=2.895$ in the COSMOS field described above.

In contrast to those DHCs with LBG HDRs, the DHC we found at $z=3.453$
has no counterpart LBG overdensity in the present statistics, as
mentioned above. It would be exceptional and may suggest peculiar
characteristics of this cluster, i.e., the DHC of a 20 cMpc scale
discovered at $z=3.453$ may possess some mechanisms or an extraordinary
structure to preferentially generate AGNs but restrict the LBG formation. 

Here, it is noteworthy that in the composite spectrum of the 5
sight-lines in the south of the pair AGN, we could not find any
significant absorption by C~{\sc ii} (1334.53\,\AA) as
well as Si~{\sc iv} (1393.76 and 1402.77\,\AA) associated with the
$z=3.453$ DHC, although the spectral resolution of $R=180$ is not high
enough to put stringent limit on metal enrichment. On
the other hand, \cite{cuc14} reported a significant detection of
Si~{\sc iv} absorption associated with their DHC having a sharp LBG peak
at $z=2.895$ in their VIMOS survey with a similar spectral resolution of
$R=230$ as ours. Unfortunately, other metal absorption such as Si~{\sc
ii} (1260.42\,\AA), O~{\sc i} (1302.17\,\AA), and Si~{\sc ii}
(1304.37\AA) associated with the DHC at $z=3.453$ can not be
investigated in our spectroscopic analysis with the low 
spectral resolution of $R=180$, because these
wavelengths fall in rather wide masked ranges affected by night emission
lines indicated by the grey shades in Figure~12 and so on.

It would be very interesting to clarify whether the
$z=3.453$ DHC has less metals or not. Deep spectroscopic survey for LBGs
and dense H~{\sc i} gas as well as metals with higher spectral
resolution than the present study is strongly desired to confirm the
underdensity of LBGs in this region and to investigate the metal
abundance of the DHC. If the metal-poor or even metal-free nature in the
DHC has been proved in such a survey, the H~{\sc i} cluster will be recognized 
as a candidate for primordial space survived at $z\sim3.5$ \citep[e.g.,][]{fum11}. 

\subsection{LBG concentration at $z=3.69$--$3.80$ associated with an AGN and H~{\sc i} absorbers}
 
\begin{figure}
 \begin{center}
  \includegraphics[width=70mm]{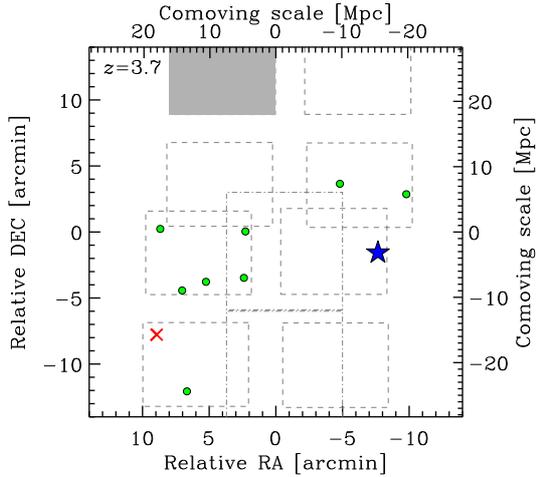}
 \end{center}
 \caption{Sky distribution of 8 LBGs (green dots) at $z=3.69$--3.80
 which form a bump in redshift distribution shown in
 Figure~\ref{zdist}~(a). The red cross indicates the sky position of the
 sight-line of the LBG at $z=4.03$, which is the most distant object in
 the present survey. The blue star shows the AGN at $z=3.801$. 
 The gray lines and shaded area are the same as in Figure~\ref{z3p10dip}.}
 \label{z3p7skymap}
\end{figure}

We can see an LBG bump consisting of 8 LBGs at $z=3.69$--3.80 in the redshift 
distribution of Figure~\ref{zdist}~(a). We show their sky map in
Figure~\ref{z3p7skymap}, where the AGN detected at  
$z=3.801$ in VI06 is also plotted with the blue star symbol and a
sight-line of the LBG found in VI08 at $z=4.03$ is indicated by the red
cross symbol. Seven out of the 8 LBGs seem to be localized in a
belt-like/filamentary structure which extends from west to east with a
mean width of about 8\,arcmin (16\,cMpc) and a length of about
20\,arcmin (40\,cMpc). We call the filamentary area a candidate HDR.  


Although the statistics is insufficient, we try to discuss the
overdensity and appearance probability of this HDR. The thickness of
this candidate HDR is estimated to be about 85 cMpc from the redshift
interval of the 7 LBGs, $dz=3.69-3.80=0.11$. Thus, the comoving volume
of the candidate HDR is $5.5 \times 10^4$ cMpc$^3$. 
When LBGs distribute uniformly in this volume, we obtain
the expected number of 2.4, assuming the selection function of the dashed
line in Figure~\ref{DAcomp}~(a). Therefore, the HDR has a number overdensity 
$\delta_{LBG}= 1.9\pm1.1$, which results in a mass
overdensity $\delta_{M}=0.73\pm0.42$, when the bias
parameter $b=2.6$ \citep{bie13,cuc14} is assumed for the LBGs. The
$1\sigma$ fluctuation of the dark matter for this volume at $z=3.75$ is
estimated to be 0.13, by using the formula in \cite{yam12}. Thus, the
candidate HDR at $z=3.7$--3.8 is also a rare event with $6\pm3\sigma$. 
%

\begin{figure}
 \begin{center}
  \includegraphics[width=70mm]{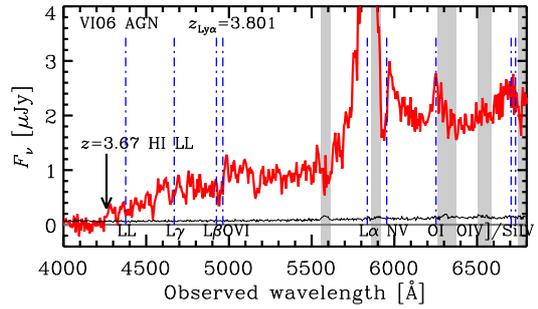} 
 \end{center}
 \caption{One-dimensional spectrum of the AGN at $z=3.801$. The LLS is
 seen at 4255~\AA\ ($z=3.67$). The gray shaded regions 
 indicate noisy wavelengths due to night emission lines. The vertical
 dot-dashed lines indicate some emission/absorption features.}       
 \label{AGN2244z3.80spec}
\end{figure}

It is interesting that the AGN with $z=3.801$ found in VI06 exists in
the envelope region of the candidate HDR as shown in
Figure~\ref{z3p7skymap}. We show the spectrum of the AGN with $R=22$ AB
in Figure~\ref{AGN2244z3.80spec}. Interestingly, we can find a probable
LLS at $z=3.67$, because of a continuum trough at wavelengths shorter
than 4255\,\AA. The following interpretation will be possible, i.e., the
candidate HDR has dense H~{\sc i} gas also in its envelope region and
the sight-line of the AGN penetrates the H~{\sc i} rich region having
the column density of about $10^{18}$ cm$^{-2}$ to make an LLS.


\begin{figure}
 \begin{center}
  \includegraphics[width=70mm]{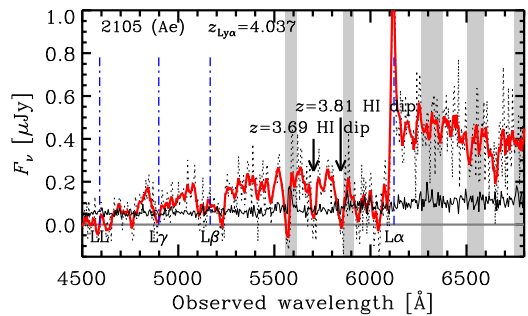}
 \end{center}
 \caption{One-dimensional spectrum of the LBG with the highest redshift
 $z=4.03$ in this survey. The gray shaded regions indicate
 noisy wavelengths due to night emission lines. The vertical dot-dashed
 lines indicate some emission/absorption features.}      
 \label{2105z4.03spec}
\end{figure}

Unlike the cases of the LSPC at $z=3.1$ and the DHC at $z=3.453$
associated with the pair AGN, there is only one LBG behind the candidate
HDR at $z=3.75$. Therefore, the composite method in this paper is not
practical to investigate H~{\sc i} absorption of the LBG HDR. However,
fortunately, a sight-line of the bright LBG of $R=24.57$ AB at $z=4.03$
found in the present survey, penetrates the envelope region of the
candidate HDR as seen in Figure~\ref{z3p7skymap}. We show the spectrum
of the LBG in Figure~\ref{2105z4.03spec}, which has two remarkable deep
absorption dips at $z=3.69$ and 3.81, if Ly$\alpha$ absorption is assumed. 
Their rest-frame equivalent widths are $-6$\,\AA\ and $-8$\,\AA\, respectively, 
indicating clusters of high column density H~{\sc i} clouds like sub-DLA/LLS. 


Although the candidate HDR at $z=3.75$ includes only 7 LBGs, it has a
$z=3.801$ AGN with a $z=3.67$ LLS and two deep H~{\sc i} absorption dips
at the HDR redshifts in the spectrum of a background LBG, in its
surrounding area. We can expect the reality of the structure. Future
surveys will reveal the interesting characteristics of this HDR.

\section{Summary and concluding remarks}

The SSA22 field has a large-scale structure of the LAEs with a comoving
volume of around $10^5$\,Mpc$^3$ at $z=3.1$. The structure contains,
besides 259 confident LAEs, 35 LABs including two gigantic ones
discovered by \cite{ste00}, which would be progenitors of the present
massive galaxies, as well as hundreds of LBGs, 
about 50 LAAs, and a number of K-band selected galaxies. So, the
structure should be called the ``large-scale proto-cluster (LSPC)''. 
For the field we carried out a spectroscopic survey for LBGs with the
VLT VIMOS, VI08 survey, and identified 78 LBGs brighter
than $R=25.4$ AB magnitude with secure spectroscopic redshifts between $z=2.5$ 
and 4, and we have obtained the redshift distribution of 171 LBGs, by combining 
ours with those of \cite{ste03}. Also, we stacked the spectra of our 
VI08 LBGs in the observer's frame by using the sophisticated method developed 
in this paper, and obtained the normalized composite spectrum. 
Analyzing these data, we have obtained the following results; 

(1) A strong H~{\sc i} absorption dip of rest-frame equivalent width of $-1.7$~\AA\ 
in the composite transmission spectrum has been found at the LSPC redshift, $z=3.1$.  
We have also found an absorption dip at $z=3.28$ with a sufficient significance. 
There seems a candidate LBG high density region (HDR) at the same redshift. The combination 
of LBG concentration and deep H~{\sc i} dip is similar to the $z=3.1$ structure. 

(2) We have detected a remarkable transparency peak at
$z=2.98$ in the composite spectrum, at which an LBG
void is found. On the other hand, we have no such peaks at $z=2.80$ and
2.89, where there are few LBGs, indicating voids or low-density regions 
(LDRs) of LBGs. In general, several voids/LDRs would be
also expected at redshifts between $z=2.55$ and 2.8, where the LBG
detection efficiency of our survey decreases. However, 
no transparency peaks are found at all at those
redshifts in spite of the considerable VLT/VIMOS sensitivities 
for the Ly$\alpha$ forest. It would mean that absence 
of or less LBGs in the ordinary void itself can not cause such a
prominent transparency peak found at $z=2.98$. 
The large mass of the 100\,cMpc away LSPC at $z=3.10$ would contribute to the 
transparency peak formation by the ``tidal force''. Such speculative 
and qualitative considerations are given in Discussion I in section 5. 

\bigskip

In the present LBG survey, VI08, we detected four AGNs
with $R<24.6$ and \cite{ste03} also found two AGNs in this field. In
addition, we identified five AGNs in our VIMOS survey carried out in 2006, VI06. 
These 11 AGNs listed in Table~7 distribute in the redshift range of $2.4<z<3.8$.  
In Discussion II in section 6, we have also investigated inhomogeneous structures 
in large scales accompanied with the AGNs at redshifts between 3.1 and 3.8, 
and the following interesting results were obtained. 

(3) The LSPC at $z=3.1$ in SSA22 shows an extremely high concentration of AGNs, i.e.,  
there exist 5 AGNs with $R<24.5$ at redshifts between $z=3.084$-3.132, where only 
0.5 AGNs are expected from the QSO/AGN LF for $z\sim3.2$ by \cite{mas12}.
In addition to them, 6 AGNs with $R>24.5$ are found at the LSPC redshift.   
These 11 AGNs in total at $z\sim3.1$ exhibit filamentary structure along the LAE density peak.  

(4) We have found two LBG HDRs associated with AGNs at $z=3.353$ and 
3.801, respectively. The former HDR consists of 19 LBGs 
and the $z=3.353$ AGN shows a ``sheet-like'' structure with the
appearance probability of $7\pm2\sigma$,   
indicating a very rare event.   
The other consists of 7 LBGs at redshifts between 3.69 and 3.801 together with the 
$z=3.801$ AGN seems to form filamentary structure. The appearance probability of this 
LBG HDR is estimated to be $6\pm3\sigma$, which also implies 
a rare event, although the statistics is insufficient. 
The HDR at $z=3.75$ including only 7 LBGs, however, has an AGN with a $z=3.67$ LLS 
of the column density of about $10^{18}$ cm$^{-2}$, and two deep H~{\sc i} absorption 
dips at the HDR redshifts in the spectrum of a background LBG, in its surrounding area. 
We can expect the reality of the structure. 

(5) Near the pair AGN at $z=3.455$, we found out a 20\,cMpc scale 
dense H~{\sc i} cluster (DHC) by detecting 
a deep dip at 5415\,\AA\ corresponding to $z=3.453$, if Ly$\alpha$
absorption is assumed, in the composite spectrum of 5 background LBGs
with the angular distances less than 10\,arcmin from the center of the
pair AGN. Also the spectrum of the AGN at $z=3.455$ seems to have 
an LLS at 4050\,\AA, i.e., $z=3.44$, and the spectrum
of the nearest sight-line from the partner AGN at $z=3.425$ shows an
absorption feature down to $z=3.40$ including the AGN redshift,
indicating that the DHC includes the pair AGN.  
Nevertheless, not only the DHC does not associate with 
any LBG HDR but also there is no LBG around it, in the present 
statistics, implying a possible underdensity of the LBGs there. 
If the presence of LBG LDR/void is proven by future observations,
the DHC will become a strange cluster showing peculiar characteristics, 
which preferentially generates AGNs but suppresses LBG formation. 
Such a region should be intensively studied. 
If the DHC shows less metal absorption or absence of it in a deep
spectroscopic survey with high spectral resolution in future, 
the H~{\sc i} cluster could become a candidate primordial space survived
at $z=3.45$.

\bigskip

As discussed so far, the SSA22 field has been found to possess a lot of
characteristic structures at several successive redshifts, $z=3.35$,
$z=3.45$ and $z=3.75$, in addition to the LSPC well established at
$z=3.1$. Each structure has a very rare appearance probability in the
present small statistics. 
If the inhomogeneous structures at $z>3.3$ are confirmed with better statistics 
in future spectroscopic surveys, as the $z=3.1$ LSPC has already been, 
the SSA22 field will become one of the key regions to test cosmology beyond 
the standard $\Lambda$CDM model, because the simultaneous appearance of such multiple 
large-$\sigma$ events including the LSPC at $z=3.1$ over a comoving Gpc scale 
would be unlikely in the standard structure formation scenario based on the 
gravitational evolution of quantum fluctuation at the inflation epoch.

\section*{Acknowledgments}

We thank Jean-Michel Deharveng and Christian Tapken for their
contribution to the VIMOS proposal, Ryosuke Yamauchi, Yuki Nakamura,
Mitsunori Horie, Eri Nakamura, for discussions and encouragements during
the VIMOS data reduction and analysis, Kazuhiro Shimasaku, Masami Ouchi,
Ikkoh Shimizu, Olivier Le F{\`e}vre, Eros Vanzella, Joe Hennawi, Monica
Turner, Gabor Worseck, and Cameron Hummels for discussions about the
results and implications.
T.H. is supported by the science research grant of Fujitok Corporation. 
A.K.I. is supported by JSPS KAKENHI Grant Number 23684010 and 26287034, 
I.I. is supported by JSPS KAKENHI Grant Number 24244018, and 
Y.M. is supported by JSPS KAKENHI Grant Number 17H04831 and 17KK0098.

\appendix

\section{Redshift catalog of LBGs in the SSA22 field}

Table~A.1 is the catalog of LBGs whose redshifts are
determined by our VLT/VIMOS observations. If $u^*$ magnitudes are
fainter than 1-$\sigma$ magnitude (27.8 AB), we list the value as upper
limits of brightness. 
These tables are published only in the on-line version. 

\begin{table*}
 \centering
 \begin{minipage}{170mm}
  \caption{Redshift catalog of LBGs in VIMOS 2008 observations.}
  \begin{tabular}{lccccccccccc}
   \hline
   & & RA & DEC & & & $u^*$ & $B$ & $V$ & $R_{\rm c}$ &
   $i'$ & $z'$ \\ 
   Slit & ID & (J2000) & (J2000) & $z_{\rm sys}$ & Category & [AB] & [AB] & [AB] &
   [AB] & [AB] & [AB] \\ 
   \hline
   1102 & 116487 & 22:18:05.97 &  0:23:14.67 & 2.921 & Ae & 26.13 & 25.29 & 24.78 & 24.75 & 24.61 & 24.73 \\
   1109 & 104357 & 22:17:42.58 &  0:20:54.73 & 3.168 & Ae & 26.16 & 24.90 & 23.98 & 24.13 & 24.01 & 24.03 \\
   1111 & 100623 & 22:17:37.14 &  0:20:04.63 & 2.732 & Ae & 26.47 & 25.76 & 25.11 & 25.19 & 24.96 & 24.87 \\
   1114 &  99220 & 22:18:04.14 &  0:19:46.88 & 3.418 & Ae & 27.50 & 26.21 & 24.91 & 24.62 & 24.36 & 24.48 \\
   1117 &  96352 & 22:18:04.33 &  0:19:17.08 & 3.013 & Ae & 26.85 & 25.67 & 25.13 & 25.05 & 24.98 & 25.08 \\
   ... \\
   \hline
  \end{tabular}\\
 \end{minipage}
\end{table*}

\label{lastpage}

\end{document}